\documentclass[letterpaper,titlepage,11pt]{article}

\pdfoutput=1

\usepackage{amssymb,amsmath,amsfonts}
\usepackage{mathrsfs,mathtools}
\usepackage{epsfig}
\usepackage[toc]{appendix}
\numberwithin{equation}{section}
\usepackage{ccaption,cite}
\usepackage{graphicx}
\usepackage{subfig}
\usepackage{tikz}
\usetikzlibrary{decorations.pathmorphing,calc}

\usepackage[
      colorlinks=true,
      linkcolor=blue,
      urlcolor=magenta,
      filecolor=green,
      citecolor=red,
      pdfstartview=FitV,
      pdftitle={},
        pdfauthor={Hamid Afshar, St\'ephane Detournay, Daniel Grumiller, Blagoje Oblak},
        pdfsubject={},
        pdfkeywords={},
        pdfpagemode=None,
        bookmarksopen=true
      ]{hyperref}

\usepackage[all]{hypcap}

\def\clock{{\count0=\time
           \divide\count0 60
           \ifnum\count0<10 0\fi\the\count0
           \multiply\count0 -60 \advance\count0 \time
           :\ifnum\count0<10 0\fi \the\count0
         }}
\newcommand{\timestamp}{{\small\vbox{\hbox{\tt\jobname.tex}
\hbox{\the\day/\the\month/\the\year, \clock}}}}

\usepackage{amsmath,amssymb}
\usepackage{verbatim}
\usepackage{graphicx}
\usepackage{hyperref}
\usepackage{color}

\DeclareFontFamily{OT1}{rsfs}{}
\DeclareFontShape{OT1}{rsfs}{m}{n}{ <-7> rsfs5 <7-10> rsfs7 <10->rsfs10}{} 
\DeclareMathAlphabet{\mycal}{OT1}{rsfs}{m}{n}

\newcommand{\Ad}{\text{Ad}}
\newcommand{\Cinf}{C^{\infty}(S^1)}
\newcommand{\Diff}{\mathrm{Diff}^+(S^1)}

\newcommand{\be}[1]{ \begin{equation}\label{#1} }
\newcommand{\ee}{\end{equation}}
\newcommand{\bea}[1]{\begin{eqnarray}\label{#1} }
\newcommand{\eea}{\end{eqnarray}}

\newcommand{\eq}[2]{\begin{equation} #1 \label{#2} \end{equation}}
\newcommand{\eps}{\varepsilon}

\newcommand{\ga}{\gamma}
\newcommand{\de}{\delta}

\DeclareMathOperator{\extdm}{d}
\newcommand{\extd}{\extdm \!}

\newcommand{\boost}{\eta} 
\newcommand{\f}{\eta} 

\begin{document}

\renewcommand{\thefootnote}{\fnsymbol{footnote}}
\begin{titlepage}
\leftline{}{TUW--15--22}
\vskip 1.5cm
\centerline{\LARGE \bf Near-Horizon Geometry and Warped Conformal Symmetry}
\vskip 1.6cm
\centerline{\bf Hamid Afshar\footnotemark[1]\footnotemark[2], St\'ephane Detournay\footnotemark[3],
Daniel Grumiller\footnotemark[4] and Blagoje Oblak\footnotemark[3]\footnotemark[5]}  
\vskip 0.5cm
\centerline{\sl \footnotemark[1]Van Swinderen Institute for Particle Physics and Gravity, University of 
Groningen} 
\centerline{\sl Nijenborgh 4, 9747 AG Groningen, The Netherlands}
\smallskip
\centerline{\sl \footnotemark[2] School of Physics, Institute for Research in Fundamental Sciences (IPM)
} 
\centerline{\sl P.O.Box 19395-5531, Tehran, Iran}
\smallskip
\centerline{\sl \footnotemark[3]Physique Th\'eorique et Math\'ematique, Universit\'e Libre de Bruxelles and 
International Solvay Institutes}
\centerline{\sl Campus Plaine C.P. 231, B-1050 Bruxelles, Belgium}
\smallskip
\centerline{\sl \footnotemark[4]Institute for Theoretical Physics, TU Wien}
\centerline{\sl Wiedner Hauptstrasse 8-10/136, A-1040 Vienna, Austria}
\smallskip
\centerline{\sl \footnotemark[5]DAMTP, Centre for Mathematical Sciences, University of Cambridge}
\centerline{\sl Wilberforce Road, Cambridge CB3 0WA, United Kingdom}

\vskip 0.5cm
\centerline{\small E-mail: \tt \href{mailto:afshar@hep.itp.tuwien.ac.at}{afshar@ipm.ir}, 
\href{mailto:sdetourn@ulb.ac.be}{sdetourn@ulb.ac.be},} 
\centerline{\small \tt \href{mailto:grumil@hep.itp.tuwien.ac.at}{grumil@hep.itp.tuwien.ac.at}, 
\href{mailto:boblak@ulb.ac.be}{boblak@ulb.ac.be}}

\vskip 1.6cm
\centerline{\bf Abstract} \vskip 0.2cm \noindent

We provide boundary conditions for three-dimensional gravity including boosted Rind\-ler 
space\-times, representing the near-horizon geometry of non-extremal black holes or flat space 
cosmologies. These boundary conditions force us to make some unusual choices, like integrating the canonical 
boundary currents over retarded time and periodically identifying the latter. The asymptotic 
symmetry algebra turns out to be a Witt algebra plus a twisted $u(1)$ current algebra with vanishing 
level, corresponding to a twisted warped CFT that is qualitatively different from the ones studied 
so far in the literature. We show that this symmetry algebra is related to BMS by a twisted 
Sugawara construction and exhibit relevant features of our theory, including matching micro- and macroscopic 
calculations 
of the entropy of zero-mode solutions. We confirm this match in a generalization to boosted Rindler-AdS. 
Finally, we show how Rindler entropy emerges in a suitable limit.

\end{titlepage}

\pagestyle{empty}
\tableofcontents
\newpage
\renewcommand{\thefootnote}{\arabic{footnote}}

\pagestyle{plain}
\setcounter{page}{1}

\newpage

\section{Introduction}\label{se:1}

Rindler space arises generically as the near horizon approximation of non-extremal black holes or 
cosmological spacetimes. Thus, if one could establish Rindler holography one may expect it to 
apply universally. In particular, obtaining a microscopic understanding of Rindler entropy could pave the way 
towards some of the unresolved puzzles in microscopic state counting, like a detailed understanding of the 
entropy of the Schwarzschild black hole, which classically is one of the simplest black holes we know, but 
quantum-mechanically seems to be among the most complicated ones. 

Our paper is motivated by this line of thought, but as we shall see the assumptions we are going to impose 
turn out to have a life of their own and will take us in somewhat unexpected directions. This is why we will label what we do in
this work as ``quasi-Rindler holography'' instead of ``Rindler holography''. 

\subsubsection*{Three-dimensional gravity}

For technical reasons we consider Einstein gravity in three spacetime dimensions 
\cite{Staruszkiewicz:1963zza}. 
While simpler than its higher-dimensional 
relatives, it is still complex enough to exhibit many of the interesting features of gravity: black holes 
\cite{Banados:1992wn
}, cosmological spacetimes \cite{Cornalba:2002fi} and boundary gravitons 
\cite{Brown:1986nw}.

In the presence of a negative cosmological constant the seminal paper by Brown and Henneaux 
\cite{Brown:1986nw} established one of the precursors of the AdS/CFT correspondence. The key ingredient to 
their discovery that AdS$_3$ Einstein gravity must be dual to a CFT$_2$ was 
the imposition of precise (asymptotically AdS) boundary conditions. This led to the realization that some of 
the bulk first class constraints become second class at the boundary, so that boundary states emerge and the 
physical Hilbert space forms a representation of the 2-dimensional conformal algebra with specific values 
for the central charges determined by the gravitational theory.

As it turned out, the Brown--Henneaux boundary conditions can be modified, in the presence of matter 
\cite{Henneaux:2002wm}, in the presence of higher derivative interactions 
\cite{Grumiller:2008es} 
and even in pure Einstein gravity \cite{Compere:2013bya}. 
In the present work we shall be concerned with a new type of boundary conditions for 3-dimensional Einstein 
gravity without cosmological constant. Let us therefore review some features of flat space Einstein gravity.

In the absence of a cosmological constant Barnich and Comp\`ere pioneered a Brown--Henneaux type of analysis 
of asymptotically flat boundary conditions \cite{Barnich:2006av} and found a specific central 
extension of the BMS$_3$ algebra \cite{Ashtekar:1996cd}, which is isomorphic to the Galilean conformal 
algebra 
\cite{Bagchi:2009my, Bagchi:2009pe, Bagchi:2010zz}. 
Based on this, there were numerous developments in the 
past few years, like the flat space chiral gravity proposal \cite{Bagchi:2012yk}, the counting of flat space 
cosmology microstates \cite{Barnich:2012xq}, 
the existence of phase transitions between flat 
space cosmologies and hot flat space \cite{Bagchi:2013lma, Detournay:2014fva},  higher spin generalizations 
of 
flat space gravity \cite{Afshar:2013vka}, 
new insights into representations and characters 
of the BMS$_3$ group \cite{Barnich:2014kra, 
Oblak:2015sea} with applications to the 1-loop 
partition function \cite{Barnich:2015mui},  flat space holographic entanglement entropy \cite{Bagchi:2014iea} 
and numerous other aspects of flat space holography \cite{Barnich:2012aw,  
Gary:2014ppa, 
Bagchi:2015wna}.

\subsubsection*{Quasi-Rindler gravity}

As we shall show in this paper, the flat space Einstein equations,
\eq{
R_{\mu\nu} = 0,
}{eq:intro1}
do allow for solutions that do not obey the Barnich--Comp\`ere boundary conditions but instead exhibit 
asymptotically Rindler behavior, 
\eq{
\extd s^2 \sim {\cal O}(r)\,\extd u^2 - 2\extd u \extd r + \extd x^2 + {\cal O}(1)\,\extd u  \extd x + \dots
}{eq:intro2}
The main goal of the present work is to set up these boundary conditions, to prove their consistency and to 
discuss some of the most relevant consequences for holography, in particular the asymptotic symmetry algebra 
and a comparison between macroscopic and microscopic entropies. Some previous literature on asymptotically 
Rindler spacetimes is provided by Refs.~\cite{Rindler:1966zz, Fulling:1972md, 
Unruh:1976db, 
Laflamme:1987ec, 
Lowe:1994ah, 
Chung:2010ge, 
Grumiller:2010bz, 
Parikh:2011aa
}. 

Before we delve into the relevant technicalities we address one conceptual issue that may appear to stop any 
attempt to Rindler holography in its track. For extremal black holes the usual near-horizon story works due 
to 
their infinite throat, which implies that one can consistently replace the region 
near the horizon by the near-horizon geometry and apply holography to the latter. By contrast, non-extremal 
black holes do not have an infinite throat. Therefore, the asymptotic region of Rindler space in general has 
nothing to do with the near-horizon region of the original theory. So even if one were to find some dual 
field 
theory in some asymptotic Rindler space, it may not be clear why the corresponding physical states should be 
associated with the original black hole or cosmological spacetime. However, as we shall see explicitly, the 
notion of a ``dual theory living at the boundary'' is misleading; one could equally say that the ``field 
theory lives near the horizon'', since (in three dimensions) the canonical charges responsible for 
the emergence of physical 
boundary states are independent of the radial coordinate. While we are going to encounter a couple 
of 
obstacles to apply Rindler holography the way we originally intended to do, we do not consider the finiteness 
of the throat of non-extremal black holes as one of them.

Starting with the assumption \eqref{eq:intro2} we are led to several consequences that we did not anticipate. 
The first one is that, on-shell, the functions specifying the space\-time metric depend on retarded 
time $u$ instead of the spatial coordinate $x$, as do the components of the asymptotic Killing vector fields. 
As a consequence, canonical charges written as integrals over $x$ all vanish and the asymptotic symmetries 
are all pure gauge. However, upon writing surface charges as integrals over $u$ and taking time to be 
periodic, the asymptotic symmetry 
algebra turns out to describe a warped CFT of a type not 
encountered before: there is no Virasoro central charge nor a $u(1)$-level; instead, there is a non-trivial 
cocycle in the mixed commutator. Based on 
these results we determine the entropy microscopically and find that it does not coincide with the naive 
Rindler entropy, as a consequence of the different roles that $u$ and $x$ play in quasi-Rindler, versus 
Rindler, holography. In the quasi-Rindler setting, we show that our microscopic result for entropy 
{\it does} match the macroscopic one. As we shall see, the same 
matching occurs in our generalization to Rindler-AdS. 

In summary, in this paper we describe a novel type of theory with interesting 
symmetries inspired by, but slightly different from, naive expectations of Rindler holography.

This work is organized as follows. In section \ref{se:2} we state boosted Rindler boundary 
conditions and provide a few consistency checks. In section \ref{se:3} we derive the asymptotic 
symmetry algebra with its central extension and discuss 
some implications for the putative dual theory. Then, in section \ref{se:4}, we address 
quasi-Rindler thermodynamics, calculate free energy and compare macroscopic 
with microscopic results for entropy, finding exact agreement between the two. Section 
\ref{se:5} is devoted to the generalization of the discussion to quasi-Rindler AdS. Finally, we 
conclude in section 
\ref{se:6} with some of the unresolved issues. Along the way, we encounter novel aspects of warped conformal 
field 
theories, which we explore in appendix \ref{AppB}. Questions related to standard Rindler thermodynamics are 
relegated to appendix \ref{app:B}.

\section{Boosted Rindler boundary conditions}\label{se:2}

The 3-dimensional line-element [$u,r,x\in(-\infty,\infty)$]
\eq{
\extd s^2 = -2a(u)\,r\,\extd u^2 - 2\extd u\extd r + 2\f(u)\,\extd u \extd x + \extd x^2
}{eq:r1}
solves the vacuum Einstein equations \eqref{eq:intro1} for all functions $a$, $\f$ that depend solely on the 
retarded time $u$. For vanishing $\f$ and constant positive $a$ the line-element \eqref{eq:r1} describes 
Rindler space with acceleration $a$. [This explains why we chose the factor $-2$ in the first term in 
\eqref{eq:r1}.] If $\f$ does not vanish we have boosted Rindler space. These observations motivate us 
to formulate consistent boundary conditions that include all line-elements of the form \eqref{eq:r1} as 
allowed classical states.

The gravity bulk action we are going to use is the Einstein--Hilbert action
\eq{
I_{\textrm{\tiny EH}} = \frac{k}{4\pi}\,\int\extd^3x\sqrt{-g}\,R\,,\qquad k=\frac{1}{4G_N}\,,
}{eq:r7}
which, up to boundary terms, is equivalent to the Chern--Simons action 
\cite{Achucarro:1987vz} 
\eq{
I_{\textrm{\tiny CS}} = \frac{k}{4\pi}\,\int\,\langle A\wedge \extd A + \tfrac23\,A\wedge A\wedge 
A\rangle
}{eq:r6}
with an $isl(2)$ connection $A$ and corresponding invariant bilinear form $\langle\cdots\rangle$. 
Explicitly, the $isl(2)$ generators span the Poincar\'e algebra in three dimensions ($n,m = 0,\pm1$),
\eq{
 [L_n,\,L_m] = (n-m)\,L_{n+m}\qquad [L_n,\,M_m]=(n-m)\,M_{n+m}\qquad [M_n,\,M_m] = 0\,,
}{eq:r5}
and their pairing with respect to the bilinear form $\langle\cdots\rangle$ is
\eq{
\langle L_1,\, M_{-1}\rangle = \langle L_{-1},\, M_1\rangle = -2 \qquad \langle L_0,\, M_0\rangle = 1
}{eq:r32}
(all bilinears not mentioned here vanish). The connection can be decomposed in components as $A = A_L^+ L_1 
+ A_L^0 L_0 + A_L^- L_{-1} + A_M^+ M_1 + A_M^0 M_0 + A_M^- M_{-1}$. In terms of these components the 
line-element reads
\eq{
\extd s^2 = g_{\mu\nu}\,\extd x^\mu \extd x^\nu = -4A_M^+ A_M^- + (A_M^0)^2\,.
}{eq:r8}
Thus, from a geometric perspective the components $A_M^n$ correspond to the dreibein and the components 
$A_L^n$ to the (dualized) spin-connection \cite{Barnich:2013yka,Afshar:2013bla}.

\subsection{Boundary conditions}

In the metric formulation, boosted Rindler boundary conditions at null infinity $r\rightarrow+\infty$ are 
given by
\eq{
g_{\mu\nu} = \begin{pmatrix}
              -2a(u,\,x)\,r + {\cal O}(1)\, & \,-1 + {\cal O}(1/r)\, & \,\f(u,\,x) + {\cal O}(1/r) \\
              g_{ru} = g_{ur} & {\cal O}(1/r^2) & {\cal O}(1/r) \\
              g_{xu} = g_{ux} & g_{xr} = g_{rx} & 1+{\cal O}(1/r)
             \end{pmatrix}
}{eq:r2}
where $a(u,x)$ and $\f(u,x)$ are arbitrary, fluctuating ${\cal O}(1)$ functions. The equations of 
motion \eqref{eq:intro1} imply homogeneity of the Rindler acceleration:
\eq{
\partial_x a(u,\,x) = 0.
}{eq:r2too}
The function $\f(u,\,x)$ and subleading terms are also constrained by the equations 
of motion. These constraints are solved by functions $\f$ that depend on $u$ only\footnote{
The most general solution of these constraints (up to gauge transformations) is actually 
$\f(u,\,x)=\f(u)+k(x)\exp{[-\int^u a(u^\prime)\,\extd u^\prime]}$. We considered this more general case 
but did not find relevant new features when $k(x)$ is non-zero. In particular, the function $k(x)$ does not 
contribute to the canonical surface charges, so we set it to zero with no loss of generality.}:
\eq{
\partial_x \f(u,\,x) = 0.
}{eq:r2tooaswell}
For simplicity, from now on we always implement the asymptotic on-shell conditions \eqref{eq:r2too}, 
\eqref{eq:r2tooaswell} together with the boundary conditions \eqref{eq:r2}, i.e., we assume both $a$ and $\f$ 
depend on the retarded time $u$ only. 

Since the $x$-independence of the functions $a$ and $\f$ has important 
consequences we stress that the conditions \eqref{eq:r2too}-\eqref{eq:r2tooaswell} are forced upon us by the 
Einstein equations and our choice of boundary conditions (\ref{eq:r2}). In fact, similar boundary conditions 
were proposed already in four dimensions \cite{Chung:2010ge}, but no attempt to identify the dual theory 
was made in that paper.

For many applications it is useful to recast these boundary conditions in first order form in terms of an 
$isl(2)$ connection
\eq{
A = b^{-1}\,\big(\extd + \frak{a} + {\cal O}(1/r)\big)\,b,
}{eq:r3}
with the ISL$(2)$ group element 
\eq{
b=\exp{\big(\tfrac r2\,M_{-1}\big)}
}{eq:r4}
and the auxiliary connection
\begin{subequations}
 \label{eq:r21}
\begin{align}
 \frak{a}_L^{+} &= 0 & \frak{a}_M^{+} &= \extd u \\
 \frak{a}_L^0 &= a(u)\,\extd u & \frak{a}_M^0 &= \extd x \\
 \frak{a}_L^{-} &= -\tfrac12\,\big(\f'(u)+a(u)\f(u)\big)\,\extd u & \frak{a}_M^{-} &= -\tfrac12\,\f(u)\,\extd 
x 
\end{align}
\end{subequations}
where $\pm$ refers to $L_{\pm1}$ and $M_{\pm1}$. Explicitly,
\eq{
A = \frak{a} + \frac{\extd r}{2}\, M_{-1} + \frac{a(u)r\,\extd u}{2} \, M_{-1}
}{eq:bigA}
where the second term comes from $b^{-1}\extd b$ and the linear term in $r$ from applying the 
Baker--Campbell--Hausdorff formula $b^{-1}\frak{a}b=\frak{a} - \frac{r}{2}\, [M_{-1},\,\frak{a}]$. Using 
(\ref{eq:r8}), this Chern--Simons connection 
is equivalent to the metric \eqref{eq:r1}.

\subsection{Variational principle}\label{se:2.2}

For a well-defined variational principle the first variation of the full action $\Gamma$ must vanish for all 
variations that preserve our boundary conditions \eqref{eq:r2}. Since we shall later employ the Euclidean 
action to determine the free energy  we use Euclidean signature here too.

We make the ansatz\footnote{
In the Lorentzian theory the boundary is time-like (space-like) if $a$ is 
positive (negative). To accommodate both signs one should replace $K$ by $|K|$. To reduce clutter 
we assume 
positive $a$ and moreover restrict to zero mode solutions, $a=\rm const.$, $\f = \rm const$. It can 
be shown that the variational principle is satisfied for non-zero mode solutions if it is satisfied 
for zero mode solutions, as long as $\partial_x g_{rx}={\cal O}(1/r^2)$. We thank Friedrich 
Sch\"oller for discussions on the variational principle and for correcting a factor two.
}
\eq{
\Gamma = -\frac{1}{16\pi G_N}\,\int\extd^3x\sqrt{g}\,R - \frac{\alpha}{8\pi G_N}\,\int\extd^2x\sqrt{\gamma}\,K
}{eq:r12}
where $\alpha$ is some real parameter. If $\alpha=1$, we recover the Gibbons--Hawking--York action 
\cite{York:1972sj}. 
If $\alpha=\tfrac 12$, we recover an action that is consistent in 
flat space holography \cite{Detournay:2014fva}. We check now which value of $\alpha$ --- if any --- is 
consistent for our boosted Rindler boundary conditions \eqref{eq:r2}.

Dropping corner terms, the first variation of the action \eqref{eq:r12} reads on-shell 
\cite{Detournay:2014fva}
\eq{
\delta \Gamma \big|_{\textrm{\tiny EOM}} = \frac{1}{16\pi G_N}\,\int\extd^2x\sqrt{\gamma}\,\big( 
K^{ij}
- 
\alpha K g^{ij} + (2\alpha - 1) K n^i n^j +(1-\alpha)\gamma^{ij} n^k \nabla_k\big)\, \delta g_{ij} \,.
}{eq:r15}
We introduce now a cut-off at $r=r_c$ and place the boundary at this cut-off, with the idea of letting 
$r_c\to\infty$ at the end of our calculations. Some useful asymptotic expressions are [$n_i= 
\delta_i^r\,\frac{1}{\sqrt{2ar_c}} + {\cal O}(1/r_c^{3/2}) $ is 
the outward pointing unit normal vector, $\gamma=2ar_c + {\cal O}(1)$ the determinant of the induced metric 
at 
the boundary and $K=K^i{}_i=\sqrt{\frac {a}{2r_c}} + {\cal O}(1/r_c^{3/2})$ the trace of extrinsic curvature]:
\begin{align}
\sqrt{\ga} K^{ij} \,\de g_{ij} &= \de a + {\cal O}(1/r_c) &
\sqrt{\ga} K g^{ij}\,\de g_{ij} &= {\cal O}(1/r_c) \\
\sqrt{\ga} K n^i n^j \,\de g_{ij} &= -\de a + {\cal O}(1/r_c) &
\sqrt{\ga} \gamma^{ij} n^k \nabla_k\,\delta g_{ij} &= {\cal O}(1/r_c)\,.
\end{align}
Inserting these expressions into \eqref{eq:r15} establishes
\eq{
\delta \Gamma \big|_{\textrm{\tiny EOM}} = \frac{1}{8\pi G_N}\,\int\extd^2x\,\big(1-\alpha\big)\,\de a + 
{\cal 
O}(1/r_c)\,.
}{eq:r16}
Therefore, picking $\alpha=1$ we have a well-defined variational principle.

Demanding a well-defined variational principle for the first order  action \eqref{eq:r6} with the 
boundary conditions \eqref{eq:r3}-\eqref{eq:r21} also requires the addition of a boundary term of the form 
\eq{
\Gamma_{\textrm{\tiny CS}} = I_{\textrm{\tiny CS}} \pm \frac{k}{4\pi}\,\int\extd u\extd x\,\langle A_u
A_x\rangle
}{eq:r49}
where the sign depends on the conventions for the boundary volume form, $\epsilon^{ux}=\mp 1$. This
result agrees with the general expression found in three-dimensional gravity in flat 
space \cite{Barnich:2013yka,Gary:2012ms}. 

In conclusion, the full Euclidean second order action suitable for quasi-Rindler holography is given by
\eq{
\Gamma = -\frac{1}{16\pi G_N}\,\int\extd^3x\sqrt{g}\,R - \frac{1}{8\pi G_N}\,\int\extd^2x\sqrt{\gamma}\,K
}{eq:r17}
where the boundary contribution is the Gibbons--Hawking--York boundary term. This action is the 
basis for Rindler thermodynamics discussed in section \ref{se:4}.

\subsection{Asymptotic symmetry transformations}

The allowed diffeomorphisms preserving the boundary conditions (\ref{eq:r2}) are generated by vector fields 
$\xi$ whose components are
\begin{subequations}
 \label{eq:r11}
\begin{align}
 \xi^u &= t(u) + {\cal O}(1/r), \\
 \xi^r &= -r\,t'(u) + {\cal O}(1), \\
 \xi^x &= p(u) + {\cal O}(1/r),
\end{align}
\end{subequations}
where $t(u)$ and $p(u)$ are arbitrary real functions. Infinitesimally, the corresponding diffeomorphisms 
take the form
\begin{equation}
u\mapsto u+\epsilon\, t(u),\qquad x\mapsto x+\epsilon\, p(u)
\label{infinitesimal}
\end{equation}
on the space\-time boundary at infinity. In other words, the gravitational system defined by the boundary 
conditions (\ref{eq:r2}) is invariant under time reparametrizations generated by 
$t(u)\partial_u-rt'(u)\partial_r$ and under time-dependent translations of $x$ generated by $p(u)\partial_x$. 
These symmetries are reminiscent of those of two-dimensional Galilean conformal field theories 
\cite{Bagchi:2009pe}.

The Lie bracket algebra of allowed diffeomorphisms (\ref{eq:r11}) is the semi-direct sum of a Witt 
algebra and a $u(1)$ current algebra. This can be seen, for instance, by thinking of $u$ as a complex 
coordinate (which will indeed be appropriate for thermodynamical applications) and expanding $t(u)$ and 
$p(u)$ in Laurent modes. Another way to obtain the same result is to take $u$ periodic, say
\eq{
u \sim u + 2\pi\,L\,
}{eq:u}
(where $L$ is some length scale), and to expand the functions $t(u)$ and $p(u)$ in Fourier modes. Introducing 
the generators
\eq{
t_n\equiv\left.\xi\right|_{t(u)=L\,e^{inu/L},\;p(u)=0}
\quad\text{and}\quad
p_n\equiv\left.\xi\right|_{t(u)=0,\;p(u)=L\,e^{inu/L}}
}{eq:r9}
then yields the Lie brackets
\begin{subequations}
 \label{eq:r10}
\begin{align}
i[t_n,\,t_m] &= (n-m)\,t_{n+m}\,, \\
i[t_n,\,p_m] &= -m\,p_{n+m}\,, \\
i[p_n,\,p_m] &= 0 \,,
\end{align}
\end{subequations}
up to subleading corrections that vanish in the limit $r\rightarrow+\infty$. Thus, quasi-Rindler boundary 
conditions differ qualitatively from usual AdS holography, which relies on conformal symmetry, and from flat 
space holography \cite{Brown:1986nw,Maldacena:1997re}, which relies on BMS 
symmetry \cite{Barnich:2006av,Bagchi:2010zz}. Instead, if there exists a dual theory for quasi-Rindler 
boundary conditions, it should be a warped CFT \cite{Detournay:2012pc} whose conformal symmetry is replaced 
by 
time-reparametrization 
invariance.\footnote{
Warped CFT symmetry algebras have appeared in the context of Topologically Massive Gravity
\cite{Compere:2008cv} (see also \cite{Compere:2009zj}), 
Lobachevsky holography \cite{Bertin:2012qw}, conformal gravity  with generalized AdS or flat boundary conditions 
\cite{Afshar:2011yh, 
Afshar:2013bla}, Lower Spin Gravity \cite{Hofman:2014loa}, and Einstein 
gravity \cite{Compere:2013bya}. 
On the field theory side these symmetries were shown to be a 
consequence, under certain conditions, of translation and chiral scale invariance \cite{Hofman:2011zj}.} We 
will return to the interpretation of this symmetry in section \ref{se:3}.

The allowed diffeomorphisms (\ref{eq:r11}) can also be obtained from the Chern-Simons formulation: 
upon looking for $isl(2)$ gauge parameters $\widehat\eps$ that obey
\eq{
\delta_{\widehat\eps} A = \extd\widehat\eps + [A,\,\widehat\eps] = {\cal O}(\delta A)
}{eq:r19}
where $\delta A$ denotes the fluctuations allowed by the boundary conditions \eqref{eq:r3}-\eqref{eq:r21}, 
and writing
\eq{
\widehat \eps = b^{-1}\,\big(\eps + {\cal O}(1/r)\big) b
}{eq:r18}
[in terms of the group element \eqref{eq:r4}], one finds
\begin{align}\label{eq:r20}
\eps =t(u)\,M_1 + p(u)\,M_0 + \Upsilon(u)\,M_{-1} +\Big( a(u)t(u) - t'(u)\Big)\,L_0\;\;\;\; 
\nonumber\\- \tfrac12\Big(\left(\f'(u)+a(u)\f(u)\right)t(u)+p'(u)\Big)\,L_{-1}\,.
\end{align}
Here the functions $t(u)$ and $p(u)$ are those of (\ref{eq:r11}), while $\Upsilon(u)$ solves the 
differential equation
\eq{ 2\Upsilon'(u)+2\Upsilon(u) a(u)+p(u) \big(\f^\prime(u) + a(u) \f (u)  \big)  =0\,.}
{ODE}
Upon imposing periodicity (\ref{eq:u}), this solution is unique.

Using 
(\ref{eq:r19})-(\ref{eq:r20}) and the on-shell connection (\ref{eq:r21}), we find that the functions $a$ and 
$\f$ transform as follows under allowed diffeomorphisms:
\begin{equation}
\label{eq:r28}
\delta a = t a' + t' a - t''
\qquad
\delta \f = t \f' + t' \f +p'\,.
\end{equation}
(Prime denotes differentiation with respect to $u$.) Note in particular that translations by $p$ 
leave $a$ invariant; note also the inhomogeneous term $t''$ 
in the infinitesimal transformation law of $a$ under conformal transformations, which hints that the 
asymptotic symmetry algebra has a central extension. We are now going to verify this.

\section{Asymptotic symmetry group}\label{se:3}

This section is devoted to the surface charges associated with the asymptotic symmetries 
(\ref{eq:r11}). First we show that the conventional approach leads to a trivial theory where all asymptotic 
symmetries are gauge transformations, as any on-shell metric is gauge-equivalent to Minkowski space. We then 
opt in subsection \ref{ssec3.2} for a non-standard definition of 
surface charges, providing us with a centrally extended asymptotic 
symmetry algebra. In subsection \ref{ssec3.3} we work out the finite transformations of the dual 
energy-momentum tensor. 

\subsection{An empty theory}

We saw in the previous section that asymptotic symmetries include time reparametrizations. This is a somewhat 
ambiguous situation: on the one hand, asymptotic symmetries are generally interpreted as global symmetries, 
but on the other hand, time reparametrizations are usually seen as gauge symmetries. In this subsection we 
show how the standard approach to surface charges in gravity selects the latter interpretation.

In the Chern--Simons formulation 
\cite{Barnich:2013yka, Gary:2012ms, Afshar:2013bla, Banados:1994tn, Afshar:2014rwa}, the variation of the 
canonical current $\frak{j}$ associated with an asymptotic symmetry generated by $\widehat\eps$ 
reads
\eq{
\delta \frak{j}[\eps] = \frac{k}{2\pi}\,\langle \widehat\eps ,\,\delta A\rangle = 
\frac{k}{2\pi}\,\langle 
\eps ,\,\delta\frak{a}\rangle\,,
}{eq:r23}
where $A$ and $\frak{a}$ are related by (\ref{eq:r3}) and $\langle\cdots\rangle$ denotes the invariant 
bilinear form (\ref{eq:r32}). The integral of that expression along a line or a circle at 
infinity gives the variation of the corresponding surface 
charge \cite{Regge:1974zd}. 
In the present case, the region $r\rightarrow+\infty$ is spanned by the coordinates $u$ and $x$, but 
only the latter is space-like.\footnote{Note that the radial dependence captured by the group element $b$ 
defined in \eqref{eq:r4} drops out of the canonical currents \eqref{eq:r23}. The corresponding charges can 
therefore be defined on any $r=\rm const.$ slice (including the horizon).} Accordingly, the natural 
surface charges are integrals over $x$; unfortunately 
the boundary conditions \eqref{eq:r21} [and the ensuing asymptotic symmetries \eqref{eq:r20}] set to zero the 
$x$-component of the variation (\ref{eq:r23}), so that these charges all vanish. It follows that, from this 
viewpoint, {\it all} asymptotic symmetries are in fact gauge symmetries; there are no global symmetries 
whatsoever, and the theory is empty.

While this conclusion is somewhat disappointing, it does not prevent us from studying the group of gauge 
symmetries in its own right, and these considerations will in fact be useful once we turn to an alternative 
interpretation. Upon integrating the infinitesimal transformations (\ref{infinitesimal}), one 
obtains finite diffeomorphisms of the plane $\mathbb{R}^2$ (spanned by the coordinates $u$ and $x$) 
given by
\begin{equation}
u\mapsto f(u),\quad x\mapsto x+p(f(u)),
\label{tsfux}
\end{equation}
where $f$ is an orientation-preserving diffeomorphism of the real line (so that $f'(u)>0$ for all $u$), 
and $p$ is an arbitrary function. Such pairs $(f,p)$ span a group
\begin{equation}
G\equiv\text{Diff}(\mathbb{R})\ltimes\text{C}^{\infty}(\mathbb{R}),
\label{G}
\end{equation}
where the vector space $\text{C}^{\infty}(\mathbb{R})$ is seen as an Abelian group with 
respect to pointwise addition. Diffeomorphisms act on functions according to
\begin{equation}
\label{aact}
\left(\sigma_f\,p\right)(u)\equiv p(f^{-1}(u)),
\quad\text{i.e.}\quad
\sigma_f\,p\equiv p\circ f^{-1},
\end{equation}
so $G$ is a semi-direct product with a group operation
\begin{equation}
\label{gggroup}
(f_1,p_1)\cdot(f_2,p_2)\equiv
\big(f_1\circ f_2,p_1+\sigma_{f_1}p_2\big).
\end{equation}
It is a centerless version of the symmetry group of warped conformal field theories.

One may then ask how finite gauge transformations affect the on-shell metrics (\ref{eq:r1}), given that the 
infinitesimal transformations are (\ref{eq:r28}). We show in appendix \ref{AppA} that, under the action of a 
gauge 
transformation $(f,p)$, the functions $\f(u)$ and $a(u)$ are mapped on new functions $\tilde{\f}$ 
and $\tilde{a}$ given by
\begin{equation}
\label{3.15}
\tilde{\f}\big(f(u)\big)
=
\frac{1}{f'(u)}\Big[\f(u)-(p\circ f)'(u)\Big],
\quad
\tilde{a}\big(f(u)\big)
=
\frac{1}{f'(u)}\left[
a(u)+\frac{f''(u)}{f'(u)}
\right]\,.
\end{equation}
It is easily verified that these transformations reduce to (\ref{eq:r28}) upon taking $f(u)=u+\epsilon t(u)$, 
replacing $p(u)$ by $\epsilon p(u)$ and expanding to first order in $\epsilon$. This formula shows explicitly 
that the phase space of the theory is empty, as any diffeomorphism $f$ such that
\begin{equation}
\label{s7q}
f'(u)=C\,\exp\Bigg[-\int\limits_0^ua(v)\extd v\Bigg]\,,\qquad C>0
\end{equation}
maps $a(u)$ on $\tilde{a}=0$. When combining this map with a suitable translation $p(u)$, the whole metric 
(\ref{eq:r1}) is mapped on that of Minkoswki space, so that indeed any solution is pure gauge. Note that the 
inhomogeneous term proportional to $f''/f'$ in the transformation law of $a$ is crucial in order for the 
latter statement  to be true.

In principle one can impose suitable fall-off conditions on the functions $a(u)$ and $\f(u)$ at future and 
past infinity, and study the subgroup of (\ref{G}) that preserves these conditions. For example, 
$a(u)\sim\text{const.}+{\cal O}(1/|u|)$ would include Rindler space\-time, potentially leading to interesting 
asymptotic symmetries at $|u|\rightarrow+\infty$. We will not follow this approach here; instead, we will try 
to make the theory non-trivial by using an unconventional prescription for the definition of surface charges.

\subsection{Quasi-Rindler currents and charges}\label{ssec3.2}

In the previous subsection we interpreted asymptotic symmetries as gauge symmetries, in accordance with the 
fact that all surface charges written as integrals over a space-like slice at infinity vanish. However, 
another interpretation is available: instead of integrating (\ref{eq:r23}) over $x$, we may decide to 
integrate it over retarded time $u$. Despite clashing with the usual Hamiltonian formalism, this approach is 
indeed the most natural one suggested by the $u$-dependent asymptotic Killing vector fields (\ref{eq:r11}) 
and the solutions (\ref{eq:r1}).

For convenience we will also assume that the coordinate $u$ is periodic as in (\ref{eq:u}). This condition 
introduces closed time-like curves and breaks Poincar\'e symmetry (even when $a=0$\,!); it sets off our 
departure from the world of 
Rindler to that of {\it quasi}-Rindler holography. While it seems unnatural from a gravitational/space\-time 
perspective, this choice is naturally suggested by our asymptotic symmetries and our phase space. In the 
remainder of this paper we explore its consequences, assuming in particular that the functions $t(u)$, 
$p(u)$, $\f(u)$ and $a(u)$ are $2\pi L$-periodic. This will in fact lead us to study new aspects of warped 
conformal field theories, which we believe are interesting in their own right.

In the quasi-Rindler case, the variation of the surface charge associated with the symmetry 
transformation $(t,p)$, evaluated on the metric $(a,\f)$, reads
\eq{
\delta Q_{(a,\f)}[t,p] = \int\limits_0^{2\pi L}\extd u\, \de\,\frak{j}_u\,.
}{eq:r24}
Using (\ref{eq:r23}) and inserting expressions \eqref{eq:r21} and \eqref{eq:r20} yields
\eq{
\delta Q_{(a,\f)}[t,p] = \frac{k}{2\pi} \, \int\limits_0^{2\pi L} \extd u\,\big(t(u)\,\de T(u) + 
p(u)\,\de 
P(u)\big),
}{eq:r25}
where
\eq{
T(u) = \f'(u) + a(u)\f(u), \qquad P(u) = a(u) \,,
}{eq:r26}
so that the charges are finite and integrable:
\eq{
Q_{(a,\f)}[t,p] = \frac{k}{2\pi} \, \int\limits_0^{2\pi L}\extd u\,\big(t(u)\, T(u) + p(u)\, 
P(u)\big)\,. 
}{eq:r27}
This expression shows in particular that the space of solutions $(a,\f)$ is dual to the 
asymptotic symmetry 
algebra\footnote{Note that the change of variables (\ref{eq:r26}) is invertible for functions on the circle: 
given the functions $a$ and 
$\f$, Eq.~(\ref{eq:r26}) specifies $T$ and $P$ uniquely; conversely, given some functions $T$ and $P$, the 
functions $a$ and $\f$ ensuring that (\ref{eq:r26}) holds are unique provided one imposes $2\pi 
L$-periodicity of the coordinate $u$.} \cite{Garbarz:2014kaa,1403.3835,Barnich:2014kra,Compere:2015knw}. More 
precisely, the 
pair of functions $(T(u),P(u))$ transforms under the coadjoint 
representation of the asymptotic symmetry group, with $T(u)$ dual to time reparametrizations and 
$P(u)$ dual to translations of $x$. This observation will be crucial when determining the 
transformation law of $T(u)$ and $P(u)$ under finite asymptotic symmetry transformations, which in turn will 
lead to a Cardy-like entropy formula.

From the variations \eqref{eq:r28} of the functions $(a,\f)$, we deduce corresponding 
variations of 
the functions $T$ and $P$ in (\ref{eq:r26}):
\begin{equation}
 \label{eq:r29}
 \delta_{(t,p)} P= t P' + t' P - t'',
 \quad
 \delta_{(t,p)} T = t T' + 2t' T+p'P+p''.
\end{equation}
This result contains all the information about the surface charge algebra, including its central 
extensions. On account of $2\pi L$-periodicity in the retarded time $u$, we can introduce the 
Fourier-mode generators
\begin{equation}
\label{TnPn}
T_n\equiv
\frac{kL}{2\pi}
\int\limits_0^{2\pi L}\extd u\,e^{inu/L}\,T(u)
\qquad
P_n\equiv
\frac{k}{2\pi}
\int\limits_0^{2\pi L}\extd u\,e^{inu/L}\,P(u)\,,
\end{equation}
whose Poisson brackets, defined by $[Q_{\xi},Q_{\zeta}]=-\delta_{\xi}Q_{\zeta}$, read
\begin{subequations}
 \label{eq:r30}
 \begin{align}
  i[T_n,\,T_m] &= (n-m)\,T_{n+m}\,, \\
  \label{eq:r30b}
  i[T_n,\,P_m] &= -m\,P_{n+m} - i\kappa\,n^2\,\de_{n+m,0}\,, \\
  i[P_n,\,P_m] &= 0\,,
 \end{align}
\end{subequations}
with $\kappa=k$. As it must be for a consistent theory, this algebra coincides with the Lie bracket 
algebra \eqref{eq:r10} of allowed diffeomorphisms, up to central extensions. Note that $T_0/L$, being the 
charge that generates time 
translations $u\mapsto u+\text{const}$, should be interpreted as the Hamiltonian, while $P_0$ is the momentum 
operator (it generates translations $x\mapsto x+\text{const}$). The only central extension in \eqref{eq:r30} 
is a twist term in the mixed commutator; it is a non-trivial 2-cocycle \cite{Unterberger:2011yya}. In 
particular, it cannot be removed by redefinitions of generators since the $u(1)$ current algebra has a 
vanishing level. 

\subsection{Warped Virasoro group and coadjoint representation}
\label{ssec3.3}

As a preliminary step towards the computation of quasi-Rindler entropy, we now 
work 
out the finite transformation laws of the functions $T$ and $P$. For the sake of generality, we will display 
the result for arbitrary central extensions of the warped Virasoro group, including a Virasoro central 
charge and a $u(1)$ level. We will use a $2\pi L$-periodic coordinate $u$, but our end result 
(\ref{tsfT})-(\ref{tsfJ}) actually holds independently of that assumption.

\subsubsection*{Finite transformations of the stress tensor}

The asymptotic symmetry group for quasi-Rindler holography is (\ref{G}) with $\mathbb{R}$ replaced by 
$S^1$, and it consists of pairs $(f,p)$, where $p$ is an arbitrary function on the circle and $f$ is a 
diffeomorphism of the circle; in particular,
\begin{equation}
\label{Fcirc}
f(u+2\pi L)=f(u)+2\pi L.
\end{equation}
(For instance, the diffeomorphisms defined by (\ref{s7q}) are generally forbidden once that 
condition is imposed.) However, in order to accommodate for inhomogeneous terms such as those appearing in 
the 
infinitesimal transformations (\ref{eq:r28}), we actually need to study the central extension of this group. 
We will call this central extension the {\it warped Virasoro group}, and we will denote it by $\hat{G}$. Its 
Lie algebra reads
\begin{subequations}
 \label{Star}
 \begin{align}
  i[T_n,\,T_m] &= (n-m)\,T_{n+m}+\frac{c}{12}\,n^3\,\delta_{n+m,0} \\
  i[T_n,\,P_m] &= -m\,P_{n+m} - i\kappa\,n^2\,\de_{n+m,0} \\
  i[P_n,\,P_m] &= K\,n\,\delta_{n+m,0}\,,
 \end{align}
\end{subequations}
and is thus an extension of (\ref{eq:r30}) with a Virasoro central charge $c$ and a $u(1)$ level $K$. Note 
that, when $K\neq 0$, the central term in the mixed commutator $[T,P]$ can be removed by defining ${\cal 
L}_n\equiv T_n+\frac{i\kappa}{K}nP_n$. In terms of generators ${\cal L}_n$ and $P_n$, the 
algebra takes 
the form (\ref{Star}) without central term in the mixed bracket, and with a new Virasoro central charge 
$c'=c-12\kappa^2/K$. [We will see an illustration of this in eq.~(\ref{eq:rads10}), in the context of 
quasi-Rindler 
gravity in AdS$_3$.] But when $K=0$ as in (\ref{eq:r30}), there is no such redefinition.

We relegate to appendix \ref{AppA} the exact definition of the warped Virasoro group $\hat G$, together with 
computations related to its adjoint and coadjoint representations. Here we simply state the result that is 
important for us, namely the finite transformation law of the stress tensor $T$ and the $u(1)$ current $P$. 
By construction, these transformations coincide with the coadjoint representation of $\hat G$, written in 
eqs.~(\ref{AtsfT})-(\ref{AtsfJ}); thus, under a finite transformation (\ref{tsfux}), the pair $(T,P)$ is 
mapped to a new pair $(\tilde T,\tilde P)$ with
\begin{eqnarray}
\tilde T(f(u))
& = &
\frac{1}{(f'(u))^2}
\bigg[
T(u)+\frac{c}{12k}\{f;u\}-P(u)(p\circ f)'(u)\nonumber\\
\label{tsfT}
&   & \qquad \qquad \quad -\frac{\kappa}{k}(p\circ f)''(u)+\frac{K}{2k}((p\circ f)'(u))^2
\bigg]\\
\label{tsfJ}
\tilde P(f(u))
& = &
\frac{1}{f'(u)}
\left[
P(u)+\frac{\kappa}{k}\frac{f''(u)}{f'(u)}-\frac{K}{k}(p\circ f)'(u)
\right],
\end{eqnarray}
where $\{f;u\}$ is the Schwarzian derivative (\ref{Schw}) of $f$. These transformations extend those of a 
standard warped CFT \cite{Detournay:2012pc}, which are recovered for $\kappa=0$. In the case 
of 
quasi-Rindler space\-times, we have $c=K=0$ and $\kappa$ is non-zero, leading to 
\begin{align}
\label{3.19}
\tilde T(f(u)) &= \frac{1}{(f'(u))^2}\left[T(u)-P(u)(p\circ f)'(u)-\frac{\kappa}{k}(p\circ f)''(u)\right]\\
\tilde P(f(u)) & = \frac{1}{f'(u)}\left[P(u)+\frac{\kappa}{k}\frac{f''(u)}{f'(u)}\right]
\label{eq:3.20}
\end{align}
which (for $\kappa=k$) actually follows from (\ref{3.15}) and the definition (\ref{eq:r26}). Note 
that 
these formulas are valid regardless of whether $u$ is periodic or not! In the latter case, $f(u)$ is a 
diffeomorphism of the real line.

\subsubsection*{Modified Sugawara construction}

Before going further, we note the following: since $P(u)$ is a Kac-Moody current, one expects that a 
(possibly modified) Sugawara construction might convert some quadratic 
combination of $P$'s into a CFT stress tensor. This expectation is compatible with the fact that 
$P(u)\extd u$ is a one-form, so that $P(u)\extd u\otimes P(u)\extd u$ is a quadratic 
density. Let us therefore define
\begin{equation}
\mathcal M(u)
\equiv
\frac{k^2}{2\kappa}\left(P(u)\right)^2+k\,P'(u)
\label{Suga}
\end{equation}
and ask how it transforms under the action of $(f,p)$, given that the transformation law of $P(u)$ is 
(\ref{eq:3.20}). Writing this transformation as $\mathcal M\longmapsto\widetilde{\mathcal M}$, the 
result is
\begin{equation}
\label{tsfpBis}
\widetilde{\mathcal M}(f(u))
=
\frac{1}{(f'(u))^2}
\Big[\mathcal M(u)
+
\kappa\,\{f;u\}
\Big],
\end{equation}
which is the transformation law of a CFT stress tensor with 
central charge
\begin{equation}
\label{eq:r39}
c_M\equiv12\kappa=12k=\frac{3}{G_N}\,.
\end{equation}
Once more, this observation is independent of whether the coordinate $u$ is periodic or not.

This construction is at the core of a simple relation between the quasi-Rindler symmetry algebra 
(\ref{eq:r30}) and the BMS$_3$ algebra. Indeed, by quadratically 
recombining the generators $P_n$ thanks to a twisted Sugawara construction
\eq{
M_n = \frac{1}{2\kappa}\,\sum_{q\in\mathbb{Z}}P_{n-q}P_q - in\,P_n  \,,
}{eq:r38}
the brackets \eqref{eq:r30} reproduce the centrally extended BMS$_3$ algebra 
\cite{Barnich:2006av}:
\begin{subequations}
\label{bmsSuga}
\begin{align}
i[T_n,\,T_m] &= (n-m)\,T_{n+m} \\
i[T_n,\,M_m] &= (n-m)\,M_{n+m} + \frac{c_M}{12}\,n^3
\,\de_{n+m,\,0}\\
i[M_n,\,M_m] &= 0.
\end{align}
\end{subequations}
This apparent coincidence 
implies that the representations of the warped Virasoro group (with vanishing Kac-Moody level) are 
related to those of the BMS$_3$ group, but one should keep in mind that the similarity of group 
structures does {\it not} imply 
similarity of the physics involved; in particular, the Hamiltonian operator in (\ref{bmsSuga}) is 
(proportional to) $T_0$, while the standard BMS$_3$ Hamiltonian is $M_0$. Nevertheless, unitary 
representations of the warped Virasoro group with vanishing Kac-Moody level $K$ can indeed be studied 
and classified along the same lines as for BMS$_3$ 
\cite{Barnich:2014kra}. 
As we do not use these results 
in the present work we relegate this discussion to appendix \ref{ureps}.

\section{Quasi-Rindler thermodynamics}\label{se:4}

In this section we study quasi-Rindler thermodynamics, both microscopically and macroscopically, assuming 
throughout that surface charges are defined as integrals over time and that the coordinate $u$ is $2\pi 
L$-periodic. We start in 
subsection \ref{se:micro} with a microscopic, Cardy-inspired derivation of the entropy of zero-mode 
solutions. Section \ref{app:A} is devoted to certain
geometric aspects of boosted Rindler spacetimes, for instance their global Killing vectors, which 
has consequences for our analytic continuation to  Euclidean signature in subsection \ref{se:4.1}. In 
subsection 
\ref{se:4.2} we evaluate the on-shell action to determine the free energy, from which we then derive other 
thermodynamic quantities of interest, such as macroscopic quasi-Rindler entropy. In particular, we exhibit 
the matching between the Cardy-based computation and the purely gravitational one.

\subsection{Modular invariance and microscopic entropy}\label{se:micro}

Here, following \cite{Detournay:2012pc}, we switch on chemical potentials (temperature and velocity) and 
compute the partition function in the high-temperature limit, assuming the validity of a suitable 
version of modular invariance. Because the $u(1)$ level vanishes in the present case, the modular 
transformations will not be anomalous, in contrast to standard behaviour in warped CFT 
\cite{Detournay:2012pc}. (This will no longer be true in AdS$_3$ --- see section \ref{se:5}.)

The grand canonical partition function of a theory at temperature $1/\beta$ and velocity 
$\boost$ is
\begin{equation}
Z(\beta,\,\boost)
=
\text{Tr}
\left(
e^{-\beta(H-\boost P)}
\right)\,,
\label{partFct}
\end{equation}
where $H$ and $P$ are the Hamiltonian and momentum operators (respectively), and the trace is taken 
over the Hilbert space of the system. In the present case, the Hamiltonian is the (quantization of the) 
charge (\ref{eq:r27}) associated with $t(u)=1$ and $p(u)=0$, i.e.~the zero-mode of $T(u)$ up to normalization:
\begin{equation}
H=\frac{k}{2\pi}\int\limits_0^{2\pi L}\extd u\,T(u)\,.
\label{eq:whyhasthisnolabel}
\end{equation}
As for the momentum operator, it is the zero-mode of $P(u)$ (again, up to normalization). If we denote by 
$I$ the Euclidean action of the system, the partition function (\ref{partFct}) can be 
computed as the integral of $e^{-I+ \boost\int\extd\tau P}$ over paths in phase space that are periodic in 
Euclidean time $\tau$ with period $\beta$. Equivalently, 
if we assume that the phase space contains one Lagrange variable at each point of space (i.e.~that we are 
dealing with a field theory), the partition function may be seen as a path integral of $e^{-I}$ over fields 
$\phi$ that satisfy $\phi(\tau+\beta,x)=\phi(\tau,x+i\beta\boost)$ since $P$ is the generator of translations 
along $x$. Note that both approaches require the combination $H-\boost P$ to be bounded from below; in 
typical 
cases (such as AdS, where $P$ is really an angular momentum operator and $\boost$ is an angular velocity), 
this is 
a restriction on the allowed velocities of the system.

Now, our goal is to find an asymptotic expression for the partition function at high temperature. To do this, 
we will devise a notion of modular invariance (actually only $S$-invariance), recalling that the symmetries 
of our theory are transformations of (the complexification of) $S^1\times\mathbb{R}$ of the form 
\eqref{tsfux}.
Seeing the partition function (\ref{partFct}) as a path integral, the variables that are integrated out live 
on a plane $\mathbb{R}^2$ spanned by coordinates $u$ and $x$ subject to 
the identifications
\begin{equation}
(u,x)\sim(u+i\beta,x-i\beta\boost)\sim(u+2\pi L,x)\,.
\label{identifications}
\end{equation}
The transformations
\begin{equation}
\tilde{u}=\frac{2\pi iL}{\beta}\,u\qquad \tilde{x}=x+\boost\,u
\label{trick}
\end{equation}
map these identifications on
\begin{equation}
(\tilde{u},\tilde{x})
\sim
(\tilde{u} - 2\pi L,\tilde{x})
\sim
\left(\tilde{u} + i\frac{(2\pi L)^2}{\beta},\tilde{x}+2\pi L\boost\right)\,.
\label{Sidentifications}
\end{equation}
While the transformations (\ref{trick}) do not belong to the group of finite asymptotic symmetry 
transformations (\ref{tsfux}), their analogues in the case of CFT's, BMS$_3$-invariant theories and warped 
CFT's \cite{Barnich:2012xq,Detournay:2012pc} apparently lead to the correct entropy formulas. 
We shall assume that the same is true here, which implies that the partition function 
$Z(\beta,\,\boost)$ satisfies a property analogous to self-reciprocity,
\begin{equation}
\label{ss17}
Z(\beta,\,\boost) = Z\left(\frac{(2\pi L)^2}{\beta},\,\frac{i\beta \boost}{2\pi L}\right)\,.
\end{equation}
This, in turn, gives the asymptotic formula
\begin{equation}
Z(\beta,\,\boost)\stackrel{\beta\rightarrow0^+}{\sim}\exp\left[-
\frac{(2\pi L)^2}{\beta}\left(H_{\text{vac}}-
\frac{i\beta \boost}{2\pi L}P_{\text{vac}}\right)\right]\,,
\label{Zaspt}
\end{equation}
where $H_{\text{vac}}$ and $P_{\text{vac}}$ are respectively the energy and momentum of the 
vacuum state. These values can be obtained from the finite transformations \eqref{3.19} [with 
$T=P=p=0$] and \eqref{eq:3.20} [with $P=0$] by considering the map $f(u)=Le^{inu/L}$ with some 
integer 
$n$ and declaring that the vacuum value of the functions $T(u)$ and $P(u)$ is zero, exactly like for the map 
between the plane and the cylinder in a CFT.  Accordingly, the vacuum values of these 
functions ``on the cylinder'', say $\tilde T_{\text{vac}}$ and $\tilde P_{\text{vac}}$, are
\begin{equation}
\label{t17}
\tilde T_{\text{vac}}
=
0
\qquad
f'(u) \tilde P_{\text{vac}} = \frac{in}{L}\cdot\frac{\kappa}{k}\,,
\end{equation}
so that $H_{\text{vac}}=0$. Choosing $|n|\neq 1$ introduces a conical excess (see subsection \ref{se:4.1}), 
or equivalently gives a map $u\mapsto L e^{inu/L}$ which is not injective, so the only 
possible choices are $n=\pm 1$. Using $P_{\text{vac}}=\tfrac{k}{2\pi}\,\int\extd f\,\tilde 
P_{\text{vac}}(f(u))=\tfrac{k}{2\pi}\,\int_0^{2\pi L}\extd u\,f'\tilde P_{\text{vac}}$ then 
establishes
\begin{equation}
P_{\text{vac}}=\pm i\kappa\quad\text{for }n=\pm 1\,.
\label{eq:r199}
\end{equation}
The asymptotic expression (\ref{Zaspt}) of the partition function thus becomes
\begin{equation}
Z(\beta,\,\boost)\stackrel{\beta\rightarrow0^+}{\sim}e^{2\pi L\kappa|\boost|}\,,
\label{eq:Z}
\end{equation}
where the sign of the dominant vacuum value in (\ref{eq:r199}) is determined by the sign of 
$\boost$. (More precisely, the vacuum $\pm i\kappa$ is selected when $\textrm{sign}(\boost)=\mp 1$.) 
The free 
energy $F\equiv -T\log Z$ is given by
\begin{equation}
F\approx -2\pi \kappa L\,|\boost|T
\end{equation}
at high temperature, and the corresponding entropy is
\begin{equation}
\label{s18}
S=-\frac{\partial F}{\partial T}\bigg|_{\boost}\approx 2\pi \kappa L|\boost|=
2\pi kL|\boost|=
\frac{2\pi L|\boost|}{4G_N}\,.
\end{equation}
In subsection \ref{se:4.2} we will see that this result exactly matches that of a macroscopic 
(i.e.~purely gravitational) computation. Before doing so, we study Euclidean quasi-Rindler 
space\-times and elucidate the origin of the vacuum configuration (\ref{t17}).

\subsection{Boosted Rindler spacetimes and their Killing vectors}\label{app:A}

Any solution of 3-dimensional Einstein gravity is locally flat and therefore locally has six Killing vector 
fields. However, these vector fields may not exist globally. We now discuss global properties of the Killing 
vectors 
of the geometry defined by the line-element \eqref{eq:r1} with the identification (\ref{eq:u}). For 
simplicity, we present our results only for 
zero-mode solutions, $a,\f=\rm const.$ 

The six local Killing vector fields are
\begin{subequations}
\begin{align}
\xi_1 &= \partial_u & \xi_4 &= e^{au}\big(\partial_u-\f\partial_x-(ar+\tfrac{1}{2}\,\f^2)\partial_r\big) \\
\xi_2 &= \partial_x &  \xi_5 &= a(\f u+x)\xi_3 - e^{-au}\partial_x\\
\xi_3 &= e^{-au}\partial_r & \xi_6 &= a(\f u+x)\xi_4 + e^{au}\big(ar+\tfrac{1}{2}\,\f^2\big)\partial_x \,.
\label{eq:app1}
\end{align}
\end{subequations}
Globally, due to our identification \eqref{eq:u}, only the Killing vectors $\xi_1$ and $\xi_2$ survive for 
generic values of $a$. The only exception arises for specific imaginary values of Rindler acceleration,
\eq{
a=\frac{in}{L} \qquad 0\neq n\in\mathbb{Z}\,,
}{eq:r103}
in which case $\xi_3$ and $\xi_4$ are globally well-defined as well. If in addition $\f=0$, then all six 
Killing vector fields can be defined globally. The non-vanishing Lie brackets between the Killing vectors are
\begin{subequations}
 \label{eq:lieKV}
\begin{align}
 & [\xi_1,\,\xi_3] = - [\xi_2,\,\xi_5] = - a \xi_3 && [\xi_1,\,\xi_5] = a\f\xi_3 - a\xi_5 \\
 & [\xi_1,\,\xi_4] = [\xi_2,\,\xi_6] = a \xi_4 && [\xi_1,\,\xi_6] = a\f\xi_4 + a\xi_6 \\
 & [\xi_3,\,\xi_6] = [\xi_4,\,\xi_5] = a \xi_2 && [\xi_5,\,\xi_6] = a\f\xi_2 - a\xi_1 \,.
\end{align}
\end{subequations}
This algebra is isomorphic to $isl(2)$, as displayed in \eqref{eq:r5}, with the identifications $M_0=\xi_2$, 
$M_+=-2\xi_4$, $M_-=-\xi_3$, $L_0=(-\xi_1+\f\xi_2)/a$, $L_+=2(\xi_6+\f\xi_4)/a$, 
$L_-=-(\xi_5+\f\xi_3)/a$. In terms of the generators $t_n$, $p_n$ of the asymptotic Lie bracket algebra 
\eqref{eq:r10} we have the identifications $t_0\sim\xi_1$, $p_0\sim\xi_2$, $t_1\sim\xi_4$ and 
$p_{-1}\sim\xi_5$; the vector field $\xi_3$ generates trivial symmetries, while the Killing vector $\xi_6$ is 
not an asymptotic Killing vector, as it is incompatible with the asymptotic behavior 
\eqref{eq:r11}. This shows in particular that the boundary conditions of quasi-Rindler gravity actually 
break Poincar\'e symmetry even when the coordinate $u$ is not periodic. Interestingly, the four 
generators $t_0$, $t_1$, $p_{-1}$, $p_0$ obey the harmonic oscillator 
algebra
\eq{
i[\alpha,\,\alpha^\dagger]=z \qquad i[H,\,\alpha]= -\alpha\qquad i[H,\,\alpha^\dagger] = \alpha^\dagger
}{eq:ho}
where the Hamiltonian is formally given by $t_0=H=\alpha^\dagger \alpha + \rm const.$, the 
annihilation/creation 
operators formally by $t_1=\alpha$, $p_{-1}=\alpha^\dagger$ and $p_0=z$ commutes with the other 
three generators. The algebra is written here in terms of Poisson brackets, but becomes the standard 
harmonic oscillator algebra after quantization. Note, however, that $t_1$ and $p_{-1}$ are not 
generally adjoint to each 
other.\footnote{More precisely, they are certainly not each other's adjoint in a unitary representation, 
although we seem to be dealing with a non-unitary representation anyway (the 
vacuum value of $P_0$ is imaginary).} In the canonical realization \eqref{eq:r30} the first commutator 
acquires an important contribution from the central extension
\eq{
i[T_1,\,P_{-1}] = P_0-i\kappa\,.
}{eq:hoho}

If we wish to identify the vacuum as the most symmetric solution then our vacuum spacetime takes the form
\eq{
\extd s^2 = -\frac{2in\,r}{L}\,\extd u^2 - 2\extd u\extd r + \extd x^2
}{eq:vac}
with some non-zero integer $n$. We shall demonstrate in subsection \ref{se:4.1} that $|n|=1$ is the only 
choice  consistent with the $u$-periodicity \eqref{eq:u}. Thus, we have uncovered yet another 
unusual feature of quasi-Rindler holography: the vacuum metric \eqref{eq:vac} with $|n|=1$ is complex. Our 
vacuum is neither flat spacetime (as one might have guessed naively) nor a specific 
Rindler spacetime, but instead it is a Rindler-like spacetime with a specific imaginary Rindler 
``acceleration'', the value of which depends on the choice of the periodicity $L$ in \eqref{eq:u}. 
Another way to see that the solution $a=\f=0$ is not maximally symmetric for finite $L$ is to 
consider the 
six local Killing vectors $\xi^{(0)}_i$ in the limit $a,\boost\to 0$:
\begin{subequations}
\begin{align}
\xi_1^{(0)} &= \partial_u & \xi_4^{(0)} &= u\partial_u-r\partial_r \\
\xi_2^{(0)} &= \partial_x & \xi_5^{(0)} &= u\partial_x+x\partial_r \\
\xi_3^{(0)} &= \partial_r & \xi_6^{(0)} &= x\partial_u+r\partial_x 
\label{eq:app2}
\end{align}
\end{subequations}
Only four of them can be defined globally, because $\xi^{(0)}_4$ and $\xi^{(0)}_5$ have a linear dependence 
on $u$ that is incompatible with finite $2\pi L$-periodicity \eqref{eq:u}.

While the vacuum metric \eqref{eq:vac} (with $n=1$) is preserved by all six Killing vectors \eqref{eq:app1}, 
at most 
three of the associated generators of the asymptotic symmetry algebra can annihilate the 
vacuum; since we expect from the discussion in subsection \ref{se:micro} that $P_0$ is non-zero we 
pick the 
vacuum by demanding that $T_0$, 
$T_1$ and $P_{-1}$ annihilate it. The generator $P_0$ then acquires a non-zero eigenvalue due to the 
central term in \eqref{eq:hoho},
\eq{
T_0|0\rangle = T_1|0\rangle = P_{-1}|0\rangle = [T_1,\,P_{-1}]|0\rangle = 0 \quad\Rightarrow\quad 
P_0|0\rangle = i\kappa |0\rangle = \frac{i}{4G_N}|0\rangle
}{eq:r144}
while the remaining two Killing vectors are excluded for reasons stated above (one acts trivially, while the 
other violates the boundary conditions for asymptotic Killing vectors). In fact, upon performing 
a shift $P_0\to P_0 - i\kappa$, the symmetry algebra (\ref{eq:r30}) becomes
\begin{subequations}
 \label{eq:conclusions}
 \begin{align}
  i[T_n,\,T_m] &= (n-m)\,T_{n+m}\,, \\
  i[T_n,\,P_m] &= -m\,P_{n+m} - i\kappa\,n(n-1)\,\de_{n+m,0}\,, \\
  i[P_n,\,P_m] &= 0\,
 \end{align}
\end{subequations}
so that the vacuum is now manifestly invariant under $T_0$, $T_1$, $P_0$ and 
$P_{-1}$.

These considerations reproduce the 
result \eqref{eq:r199} and thereby provide a consistency check. The fact that the eigenvalue of $P_0$ is 
imaginary indicates that the warped Virasoro group is represented in a non-unitary way if the assumption of 
self-reciprocity (\ref{ss17}) holds. 

\subsection{Euclidean boosted Rindler}\label{se:4.1}

In order to prepare the ground for thermodynamics, we now study the Euclidean version of the metrics 
(\ref{eq:r1}). We consider only zero-mode solutions for simplicity. Then, defining new coordinates
\eq{
\tau = u + \frac{1}{2a}\,\ln{\big(2ar+\f^2\big)}\qquad y = x - \frac{\f}{2a}\,\ln{\big(2ar+\f^2\big)}\qquad 
\rho 
= r + \frac{\f^2}{2a},
}{eq:r13}
the line-element \eqref{eq:r1} becomes
\eq{
-2a\rho\,\extd\tau^2 + \frac{\extd\rho^2}{2a\rho} + \big(\extd y + \f\extd\tau\big)^2\,.
}{eq:r101}
For non-zero Rindler acceleration, $a\neq0$, there is a Killing horizon at $\rho_h=0$, or equivalently at
\eq{
r=r_h = -\frac{\f^2}{2a} \,.
}{eq:r102}
The patch $\rho>0$ coincides with the usual Rindler patch for positive Rindler acceleration $a$. For negative 
$a$ the patches $\rho>0$ and $\rho<0$ switch their roles. We assume positive $a$ in this work so that $\tau$ 
is a timelike coordinate in the limit $r\to\infty$.

The Euclidean section of the metric (\ref{eq:r101}) is obtained by defining
\eq{
t_{\textrm{\tiny E}}=-i\tau 
}{eq:r104}
which yields the 
line-element
\eq{
\extd s^2= 2a\rho\,\extd t^2_{\textrm{\tiny E}} + \frac{\extd\rho^2}{2a\rho}+\big(\extd y+i\f\extd 
t_{\textrm{\tiny E}}\big)^2\,.
}{eq:r14}
Demanding the absence of a conical singularity at $\rho=0$ and compatibility with \eqref{eq:u} leads to the 
periodicities
\eq{
(t_{\textrm{\tiny E}},\,y)\sim(t_{\textrm{\tiny E}}+\beta,\,y-i\beta\boost)\sim(t_{\textrm{\tiny E}}-2\pi i 
L,\,y)\,.
}{eq:r33}
which are the Euclidean version of the periodicities (\ref{identifications}), with the inverse temperature 
$\beta=T^{-1}$ given by
\eq{
T = \frac{a}{2\pi}\,.
}{eq:r34}
Given the periodicities (\ref{eq:r33}), we can now ask which values of $n$ in (\ref{eq:r103}) give rise to a 
regular space\-time with metric (\ref{eq:vac}). Consider the Euclidean line-element \eqref{eq:r14} and define 
another analytic continuation,
\eq{
a = \frac{in}{L}\qquad \tau = i t_{\textrm{\tiny E}} \qquad \hat\rho = -i\rho\,\textrm{sign}(n)\,,
}{eq:anacon}
which yields
\eq{
\extd s^2= \frac{2|n|\hat\rho\extd\tau^2}{L} + \frac{L\extd\hat\rho^2}{2|n|\hat\rho}+\big(\extd 
y+\f\extd\tau\big)^2\,.
}{eq:r110}
with the periodicities
\eq{
(\tau,\,y)\sim(\tau+i\beta,\,y-i\beta\boost)\sim(\tau+2\pi L,\,y)\,.
}{eq:r111}
The point now is that the Euclidean line-element \eqref{eq:r110} with the periodicities \eqref{eq:r111} has a 
conical singularity at $\hat\rho=0$ unless $|n|=1$. Thus, we conclude that the vacuum spacetime (in the 
sense of being singularity-free and maximally symmetric) is given by \eqref{eq:vac} with $|n|=1$, confirming 
our discussion in sections \ref{se:micro} and \ref{app:A}.

\subsection{Macroscopic free energy and entropy}\label{se:4.2}

The saddle point approximation of the Euclidean path integral leads to the Euclidean partition function, 
which in turn yields the free energy. The latter is given by temperature times the Euclidean action 
\eqref{eq:r17} evaluated on-shell:
\eq{
F = -T\,\frac{1}{8\pi G_N}\,\int\limits_0^{-i2\pi L}\extd t_{\textrm{\tiny 
E}}\int\limits_0^{i\beta|\boost|}\extd y\,\sqrt{\gamma}K\big|_{\rho\to\infty}
=  -\frac{2\pi L|\boost| T}{4G_N}\,,
}{eq:r51}
where we have inserted the periodicities from subsection \ref{se:4.1} and used $\sqrt{\gamma}K=a + {\cal 
O}(1/\rho)$. The absolute value for $\boost$ was introduced in order to ensure a positive volume 
form\footnote{
One pragmatic way to get the correct factors of $i$ is to insert the Euclidean periodicities in the ranges of 
the integrals and to demand again positive volume when integrating the function $1$. In the flat space 
calculation this implies integrating $u$ from $0$ to $\beta$, while here it implies integrating $u$ from $0$ 
to $-i2\pi L$.
} for positive $L$ and $\beta$.

From this result we extract the entropy
\eq{
S = -\frac{\partial F}{\partial T}\bigg|_{\boost} = \frac{2\pi L|\boost|}{4G_N} \,,
}{eq:r52}
which coincides with the Cardy-based result \eqref{s18}.
As a cross-check, we derive the same expression in the Chern--Simons formulation by analogy to the flat space 
results \cite{Gary:2014ppa}: 
\eq{
S = \frac{k}{2\pi}\,\int\limits_0^{-i2\pi L}\extd u\int\limits_0^{i\beta|\boost|}\extd x \,\langle A_u 
A_x\rangle  =  k\,L\,\beta|\boost|\,\langle \frak{a}_u \frak{a}_x\rangle =  2\pi k\,L |\boost| =  \frac{2\pi 
L|\boost|}{4G_N}.
}{eq:r37}
We show in the next section that the same matching occurs in Rindler-AdS.

\section{Boosted Rindler-AdS}\label{se:5}

In this section we generalize the discussion of the previous pages to the case of Rindler-AdS 
spacetimes. In subsection \ref{se:5.1} we establish quasi-Rindler-AdS boundary conditions and show 
that the asymptotic 
symmetry algebra can be untwisted to yield a standard warped CFT algebra, with a $u(1)$ level that 
vanishes in the limit of infinite AdS radius. Then, in subsection \ref{se:5.2} we derive the 
entropy microscopically, and we show in subsection \ref{se:5.3} that the same result can be 
obtained macroscopically.

\subsection{Boundary conditions and symmetry algebra}\label{se:5.1}

We can deform the metric \eqref{eq:r1} to obtain a solution of Einstein's equations $R_{\mu\nu} - 
\tfrac12\,g_{\mu\nu} R =
g_{\mu\nu}/\ell^2$ with a negative 
cosmological constant $\Lambda = -1/\ell^2$,
\eq{\extd s^2=-2a(u)r\,\extd u^2-2\extd u\extd r+2(\f(u)+2r/\ell)\extd u\extd x + \extd x^2\,.}{eq:rads1}
Starting from this ansatz, we now adapt our earlier discussion to the case of a non-vanishing cosmological 
constant. Since the computations are very similar to those of the quasi-Rindler case, we will simply 
point 
out the changes that arise due to the finite AdS radius. When we do not mention a result 
explicitly we imply that it is the same as for the flat configuration; in particular we assume 
again $2\pi L$-periodicity in $u$.

The Chern--Simons formulation is based on the deformation of the $isl(2)$ algebra to $so(2,2)$, where the 
translation generators no longer commute so that the last bracket in \eqref{eq:r5} is replaced by
\begin{align}\label{eq:rads2}
[M_n,\,M_m] = \frac{1}{\ell^2}\,(n-m)L_{n+m}\,.
\end{align}
The on-shell connection \eqref{eq:r3}-\eqref{eq:r21} and the asymptotic symmetry 
generators \eqref{eq:r18}-\eqref{ODE} are modified as
\eq{
\frak{a}\rightarrow\frak{a}+\Delta \frak{a}/\ell\,,\qquad\varepsilon\rightarrow\varepsilon+\Delta 
\varepsilon/\ell \qquad\text{and}\qquad b=\exp{\big[\tfrac r2\left(M_{-1}-\tfrac1\ell L_{-1}\right)\big]}
}{eq:rads3}
where 
\begin{subequations}
\label{eq:rads4}
\begin{align}\label{eq:radsDeltaa}
\Delta \frak{a} &=\extd u\,L_{1}-\extd x\,L_0+ \tfrac{1}{2}\f(u)\extd x\,L_{-1}  \\
\Delta \varepsilon&=  t(u)L_1-p(u)L_0-\Upsilon(u)L_{-1}\,. 
\end{align}
\end{subequations}
The connection $A$ changes correspondingly as compared to \eqref{eq:bigA},
\eq{
A\to A + \Delta A/\ell \qquad \Delta A = \Delta\frak a - \frac{\extd r}{2}\,L_{-1} + 
r\,\big(\tfrac1\ell\,L_{-1}\,\extd x -  M_{-1}\,\extd x - \tfrac12\,a(u)L_{-1}\,\extd u\big)\,.
}{eq:bigAtoo}
Note in particular that all quadratic terms in $r$ cancel due to the identity $[L_{-1},\,[L_{-1},\,\frak a]] 
- 2\ell
[L_{-1},\,[M_{-1},\,\frak a]] + \ell^2 [M_{-1},\,[M_{-1},\,\frak a]] = 0$. Plugging the result 
\eqref{eq:bigAtoo} into the
line-element \eqref{eq:r8} yields
\eq{
\extd s^2 \to \extd s^2 + \Delta \extd s^2/\ell \qquad \Delta \extd s^2 = 4r\,\extd u \extd x
}{eq:changeds}
thus reproducing the solution \eqref{eq:rads1}.

Consequently, the variations of the functions $a(u)$ and $\f(u)$ in \eqref{eq:r28} are also 
modified,
\begin{align}\label{eq:rads6}
\delta a(u)&\rightarrow\delta a(u) - 2p'(u)/\ell\,,\\[.15truecm]
\delta \f(u)&\rightarrow \delta \f(u) + 2p(u)\f(u)/\ell + 4\Upsilon(u)/\ell\,.
\end{align}
Using \eqref{ODE}, one can show that the presence of the last term in the second line does not affect the 
transformation of the function $T(u)$ defined in (\ref{eq:r26}). Moreover, the charges (\ref{eq:r27}) remain 
unchanged. In fact, in the Rindler-AdS case only the transformation of the current $P$ is deformed as
\eq{
\delta_p P=-2p'/\ell \,,
}{eq:rads7}
which leads to the following Poisson brackets of the charges $P_n$ defined in (\ref{TnPn}):
\begin{align}\label{eq:rads8}
i[P_n,\,P_m] &= -\frac{2k}{\ell}\,n\, \delta_{n+m,0} \, .
\end{align}
In particular, the limit $\ell\rightarrow\infty$ reproduces the algebra \eqref{eq:r30}.

At finite $\ell$ the presence of the non-vanishing level in \eqref{eq:rads8} enables us to remove the central 
extension of the mixed bracket of (\ref{eq:r30b}) thanks to a twist
\eq{
\mathcal L_n=T_n-\frac{i\ell\kappa}{2k}nP_n
}{eq:rads10}
in terms of which the asymptotic symmetry algebra reads
\begin{subequations}
\label{eq:rads9}
\begin{align}
i[\mathcal L_n,\,\mathcal L_m] &= (n-m)\,\mathcal L_{n+m}+\frac{c}{12}\,n^3\,\delta_{n+m,0}\\
i[\mathcal L_n,\,P_m] &= -m\, P_{n+m} \\
i[P_n,\,P_m] &= -\frac{2k}{\ell}\,n\, \delta_{n+m,0} 
\end{align}
\end{subequations}
with the expected Brown--Henneaux central charge\footnote{This central charge is expected to be 
shifted quantum mechanically at finite $k\ell$ \cite{Afshar:2011yh}. Since we are interested in 
the semi-classical limit here, we shall not take such a shift into account.} \cite{Brown:1986nw}
\eq{
c=6\frac{\kappa^2}{k}\ell
=6k\ell=\frac{3\ell}{2G_N}\,.
}{eq:c}

\subsection{Microscopic quasi-Rindler-AdS entropy}\label{se:5.2}

As in the case of a vanishing cosmological constant, it is possible to derive a Cardy-like 
entropy formula that can be applied to zero-mode solutions. The only difference with respect to 
subsection 
\ref{se:micro} is the non-vanishing $u(1)$ level $K=-2k/\ell$ that leads to a slightly different 
form of modular invariance. Namely, according to \cite{Detournay:2012pc}, the self-reciprocity of 
the partition 
function, eq.~(\ref{ss17}), now becomes
\begin{equation}
\label{modK}
Z(\beta,\,\hat\boost) = e^{\beta KL\hat\boost^2/2} Z\left(\frac{(2\pi L)^2}{\beta},
\,\frac{i\beta\hat\boost}{2\pi L}\right)\,
\end{equation}
leading to the high-temperature free energy
\begin{equation}
F\approx
\frac{(2\pi L)^2}{\beta^2}\left(H_{\text{vac}}
-\frac{i\beta\hat\boost}{2\pi L}P_{\text{vac}}\right)-KL\hat\boost^2/2\,.
\end{equation}
(We have renamed the chemical potential conjugate to $P_0$ as $\hat\boost$ for reasons that will become clear 
in subsection \ref{se:5.3}.) This is the same as in (\ref{Zaspt}), up to a temperature-independent constant 
proportional to the 
$u(1)$ level. The vacuum values of the Hamiltonian and the momentum operator are once more given by 
the arguments above (\ref{t17}); in particular, the $u(1)$ level plays no role for these 
values. Accordingly, the free energy at high temperature $T=\beta^{-1}\gg 1$ boils down to
\begin{equation}
F\approx-2\pi Lk|\hat\boost|T+kL\hat\boost^2/\ell
\label{wtf}
\end{equation}
and the corresponding entropy is again given by (\ref{s18}):
\begin{equation}
\label{eq:r113}
S = \frac{2\pi L|\hat\boost|}{4G_N}
\end{equation}
Notably, this is independent of the AdS radius. The same result can be obtained by absorbing the twist 
central charge through the redefinition 
\eqref{eq:rads10} and then using the Cardy-like entropy formula derived in \cite{Detournay:2012pc}. 
In 
the next subsection we show that this result coincides with the gravitational entropy, as in the 
flat quasi-Rindler case discussed previously.

\subsection{Macroscopic quasi-Rindler-AdS entropy}\label{se:5.3}

Generalizing the macroscopic calculations from section \ref{se:4} for zero-mode solutions \eqref{eq:rads3} 
with constant $a$ and $\f$ we find that the outermost Killing horizon is located at
\eq{
r_h = \frac{\ell}{4}\, \Big(\sqrt{a^2\ell^2+4a\ell \f} - a\ell - 2\f \Big) = -\frac{\f^2}{2a} + {\cal 
O}(1/\ell)
}{eq:r115}
and has a smooth limit to the quasi-Rindler result \eqref{eq:r102} for infinite AdS radius $\ell\to\infty$.
We assume $\eta > -a\ell/4$ so that $r_h$ is real and surface-gravity is non-zero.

The vacuum spacetime reads
\eq{
\extd s^2 = -\frac{2ir\,\extd u^2}{L} - 2\extd u\extd r + \frac{4r}{\ell}\,\extd u\extd x + \extd x^2
}{eq:r124}
where again we defined ``vacuum'' as the unique spacetime compatible with our boundary conditions, regularity 
and maximal symmetry. 

Making a similar analytic continuation as in subsection \ref{se:4.1} we obtain the line-element
\eq{
\extd s^2 = K(r)\,\extd t_{\textrm{\tiny E}}^2 + \frac{\extd r^2}{K(r)} + \big(\extd y+i(2r/\ell+\f)\extd 
t_{\textrm{\tiny E}}\big)^2
}{eq:r116}
with the Killing norm 
\eq{
K(r) = \frac{4r^2}{\ell^2} + \frac{4\f r}{\ell} + 2ar + \f^2
}{eq:r117}
and the periodic identifications
\eq{
(t_{\textrm{\tiny E}},\,y)\sim(t_{\textrm{\tiny E}}-2\pi i L,\,y)\sim(t_{\textrm{\tiny 
E}}+\beta,\,y-i\beta\hat\boost)
}{eq:r118}
where inverse temperature $\beta=T^{-1}$ and boost parameter $\hat\boost$ are given by
\eq{
T = \frac{\sqrt{a^2+4a\f/\ell}}{2\pi}
\qquad
\hat\boost = \sqrt{a^2\ell^2/4+a\f\ell}-a\ell/2\,.
}{eq:r119}
In particular, the chemical potential $\hat\eta$ no longer coincides with the parameter $\eta$ 
appearing in the metric. Note that the limit of infinite AdS radius is smooth and leads to the 
expressions in subsection \ref{se:4.1} for line-element, periodicities, temperature and boost parameter.

Converting the zero-mode solution \eqref{eq:r116}-\eqref{eq:r119} into the Chern-Simons formulation and using 
formula \eqref{eq:r37} then yields the entropy
\eq{
S = \frac{2\pi L |\hat\boost|}{4G_N}\,.
}{eq:r123} 
where independence of the AdS radius follows from the fact that the 
connection given by (\ref{eq:radsDeltaa}) has no non-zero component along any of the $M_n$'s.
This agrees with the microscopic result \eqref{eq:r113}. 

The macroscopic free energy compatible with the first law $\extd F = - S \extd T-P_0\extd\hat\eta$ is given 
by
\eq{
F(T,\,\hat\eta) = H(S,\,P_0) - TS - P_0\hat\eta  = -2\pi Lk|\hat\boost|T+kL\hat\boost^2/\ell
}{eq:r133}
where $H$ is given by the zero mode charge $T=a\eta$ through \eqref{eq:whyhasthisnolabel}, $H=kLa\eta=-F$, 
and $P_0$ is given by the zero mode charge $P=a$ through the right eq.~\eqref{TnPn}, $P_0=kLa$. The 
result \eqref{eq:r133} also coincides with the microscopic one \eqref{wtf}.

\section{Rindlerwahnsinn?}\label{se:6}

In this final section\footnote{
The play on words in this section title is evident for German speaking physicists; for the remaining ones we 
point out that ``Rinderwahnsinn'' means ``mad cow disease'' and ``Wahnsinn'' means ``madness''.
} we highlight some of the unusual features that we unraveled in our quest for near horizon holography. We 
add some comments, explanations and possible resolutions of the open issues.

\subsection{Retarded choices?}\label{se:6.1}

Let us summarize and discuss aspects of the dependence of Rindler acceleration on retarded time. (Note that 
our whole paper can easily be sign-flipped to advanced time $v$, which may be useful in 
some applications.)

\paragraph{Rindler acceleration depends on retarded time.} We started with the ansatz \eqref{eq:intro2} since 
we wanted a state-dependent Rindler acceleration to accommodate a state-dependent 
temperature. We left it open whether Rindler acceleration $a$ was a function of retarded time $u$, spatial 
coordinate $x$ or both. The Einstein equations forced us to conclude that Rindler 
acceleration can depend on retarded time only. We give now a physical reason why this should be expected. 
Namely, if the zeroth law of black hole mechanics holds then surface gravity (and thus Rindler 
acceleration) must be constant along the horizon. In particular, it cannot depend on $x$. If the horizon 
changes, e.g.~due to emission of Hawking quanta or absorption of matter, then Rindler 
acceleration can change, which makes the dependence on $u$ natural, much like the corresponding dependence of 
Bondi mass on the lightlike time.

\paragraph{Retarded time is periodic.} While many of our results are actually independent of the choice 
\eqref{eq:u}, it was still a useful assumption for several purposes, e.g.~the introduction of 
Fourier modes. For some physical observables it is possible to remove this assumption by taking the limit 
$L\to\infty$. We shall provide an important example in section \ref{se:6.3} below.

\paragraph{Boundary currents are integrated over retarded time.} If we wanted our theory to be non-empty we 
could not use the standard definition of canonical charges integrated over space, but 
instead had to consider boundary currents integrated over retarded time. We have no further comments on this 
issue, except for pointing out that in four dimensions, a bilinear in the Bondi news is integrated over 
retarded time in 
order to yield the ADM mass. Thus, despite of the clash with the usual Hamilton formulation we 
believe that we have made here the most natural choice, given our starting point.

\subsection{Rindler entropy?}\label{se:6.3}

Let us finally take a step back and try to connect with our original aim of setting up Rindler holography and 
microscopically calculating the Rindler entropy \cite{Laflamme:1987ec}. We summarize in appendix \ref{app:B} 
results for Rindler thermodynamics and Rindler entropy, which match the near horizon results of BTZ black 
holes and flat space cosmologies. 

We consider a limiting procedure starting with our result for entropy \eqref{eq:r52}. We are 
interested in a limit where simultaneously the compactification length $L$ in \eqref{eq:u} tends to 
infinity, the boost parameter $\boost$ tends to zero, the length of the spatial cycle $x$ appears in the 
entropy and all unusual factors of $i$ are multiplied by something infinitesimal. In other 
words, we try to construct a limit towards Rindler entropy \eqref{eq:r35}.

Consider the identifications \eqref{identifications} with a complexified $\beta\to\beta_0+2\pi i L$ and split 
them into real and imaginary parts:
\eq{
\textrm{Re}:\;(u,x)\sim(u,x+2\pi L\boost)\sim(u+2\pi L,x)\qquad 
\textrm{Im}:\;(u,x)\sim(u+i\beta_0,x-i\beta_0\boost)
}{eq:RE1}
The rationale behind this shift is that the real part of the periodicities untwists. As in appendix 
\ref{app:B} we call the (real) length of the $x$-cycle $\tilde L$ and thus have the relation
\eq{
\tilde L = 2\pi L\boost\,.
}{eq:RE2}
Therefore, taking the decompactification limit for retarded time, $L\to\infty$, while keeping fixed $\tilde 
L$ simultaneously achieves the desired $\boost\to 0$, so that the periodicities 
\eqref{eq:RE1} in this limit simplify to
\eq{
\textrm{Re}:\;(u,x)\sim(u,x+\tilde L) \qquad 
\textrm{Im}:\;(u,x)\sim(u+i\beta_0,x)
}{eq:RE3}
which, if interpreted as independent periodicities, are standard relations for non-rotating horizons at 
inverse temperature $\beta_0$ and with a length of the spatial cycle given by $\tilde L$. Apart 
from taking limits our only manipulation was to shift the inverse temperature $\beta$ in the complex plane. 
Thus, any observable that is independent from temperature should remain unaffected by such a 
shift; moreover, the ``compactification'' of $\beta$ along the imaginary axis is then undone by taking the 
decompactification limit $L\to\infty$. We conclude from this that entropy $S$ from 
\eqref{eq:r52} should have a smooth limit under all the manipulations above and hopefully yield the Rindler 
result \eqref{eq:r35}. This is indeed the case:
\eq{
\lim\nolimits_{L\to\infty,\boost\to\tilde L/(2\pi L)}\big( \lim\nolimits_{\beta\to\beta_0 + 2\pi i L} S \big) = 
\frac{\tilde L}{4 G_N}\,.
}{eq:RE4}
Thus, we recover the usual Bekenstein--Hawking entropy law as expected from Rindler holography. In 
this work we have also provided a microscopic, Cardy-like derivation of this result. 
A different singular limit was considered in \cite{Shaghoulian:2015dwa}, where the Rindler entropy \eqref{eq:RE4} was derived from a Cardy formula for holographic hyperscaling theories.

\subsection{Other approaches}\label{se:6.4}

\paragraph{Relation to BMS/ultrarelativistic CFT.} Using our quadratic map \eqref{eq:r38} of warped
CFT generators to BMS$_3$ generators together with the result \eqref{eq:r39} for 
the central charge, we now check what microstate counting would be given by an ultrarelativistic CFT 
(or equivalently a Galilean CFT) \cite{Barnich:2012xq}. Using the 
``angular 
momentum'' $h_L=a\f$, the ``mass'' $h_M=a^2/(2k)$, the central charge $c_M=12k$ with 
$k=1/(4G_N)$, and introducing an extra factor of $L$ to accommodate our 
periodicity \eqref{eq:u}, the ultrarelativistic Cardy formula gives
\eq{
S_{\textrm{\tiny UCFT}} = 2\pi L\,|h_L| \sqrt{\frac{c_M}{24 h_M}} = 2\pi 
L\,|a\f|\sqrt{\frac{k^2}{a^2}} = 
\frac{2\pi L|\f|}{4G_N}\,.
}{eq:r50}
This entropy thus coincides with the warped CFT entropy \eqref{s18}, and matches the gravity result 
\eqref{eq:r52}.

\paragraph{Other Rindler-type boundary conditions.}
While finishing this work the paper \cite{Donnay:2015abr} appeared which proposes alternative Rindler 
boundary conditions, motivated partly by Hawking's claim 
that the information loss paradox can be 
resolved by considering the supertranslation of the horizon caused by ingoing particles 
\cite{Hawking:2015qqa}.\footnote{%
After posting this paper, a more detailed account of the relationship between near horizon properties and supertranslations was posted by Hawking, Perry and Strominger \cite{Hawking:2016msc}, where they argue that supertranslations generate ``soft hair'' on black holes, where ``soft'' means ``zero energy''.
}
(See also \cite{Blau:2015nee,Ayon-Beato:2015xsz}.) In 
\cite{Donnay:2015abr} the state-dependent functions depend on the spatial coordinate and thus 
allow for 
standard canonical charges. The corresponding Rindler acceleration (and thus temperature) is 
state-independent and the 
asymptotic symmetry algebra has no central extension. We checked that the Rindler acceleration of that 
paper can be made state-dependent, but in accordance 
with our discussion in section \ref{se:6.1} it cannot depend on the spatial coordinate; only 
dependence on retarded/advanced time is possible. Thus, we believe that if one wants to allow for a 
state-dependent temperature in Rindler holography the path described in the present work is 
unavoidable.

\paragraph{Generalizations.} We finish by mentioning a couple of interesting generalizations that 
should allow in several cases straightforward applications of our results, like generalizations to 
higher derivative theories of (massive or partially massless) gravity, theories that include matter 
and theories in higher dimensions. In particular, it would be interesting to generalize our 
discussion to topologically massive gravity \cite{Deser:1982vy} 
in order to see how 
the entropy computation would be affected.

\section*{Acknowledgments}

We are grateful to Gaston Giribet for collaboration and for sharing his insights on Rindler holography during 
the first two years of this project (August 2013 --- June 2015). In addition, we thank 
Glenn Barnich, Diego Hofman and Friedrich Sch\"oller for discussions.

HA was supported in part by the Dutch stichting voor Fundamenteel Onderzoek der Materie (FOM) and in part by 
the Iranian National Science Foundation (INSF). SD is a Research Associate of the Fonds de la Recherche 
Scientifique F.R.S.-FNRS (Belgium). He is supported in part by the ARC grant ``Holography, Gauge Theories and 
Quantum Gravity Ð Building models of quantum black holes", by IISN - Belgium (convention 4.4504.15) and 
benefited from the support of the Solvay Family. He thanks Gaston Giribet and Buenos Aires University for 
hosting him while this work was in progress. DG was supported by the Austrian Science Fund (FWF), projects 
Y~435-N16, I~952-N16, I~1030-N27 and 
P~27182-N27, and by the program Science without Borders, project CNPq-401180/2014-0. DG also thanks Buenos 
Aires University for hosting a research visit, supported by OeAD project AR 09/2013 and 
CONICET, where this work was commenced.
BO was supported by the Fund for Scientific Research-FNRS Belgium (under grant
number FC-95570) and by a research fellowship of the 
Wiener-Anspach Foundation.

\appendix

\section{On representations of the warped Virasoro group}\label{AppB}

This appendix is devoted to certain mathematical considerations regarding the warped Virasoro group. They are 
motivated by the questions encountered in section \ref{se:3} of this paper, although they 
are also interesting in their own right. First, in subsection \ref{AppA} we study the coadjoint 
representation of the warped Virasoro group, which is needed to derive formulas 
(\ref{tsfT})-(\ref{tsfJ}) for the transformation law of the stress tensor. Then, in subsection \ref{ureps} we 
classify all irreducible, unitary representations of this group with 
vanishing $u(1)$ level using the method of induced representations. We assume throughout that the coordinate 
$u$ is $2\pi L$-periodic.

\subsection{Coadjoint representation}\label{AppA}

We call {\it warped Virasoro group}, denoted $\hat G$, the general central extension of the group
\begin{equation}
\label{wVirr}
G=\text{Diff}^+(S^1)\ltimes\text{C}^{\infty}(S^1),
\end{equation}
where the notation is the same as in (\ref{G}) up to the replacement of $\mathbb{R}$ by $S^1$ [so that in 
particular $f$ satisfies property (\ref{Fcirc})]. In this 
subsection we display an explicit definition of $\hat G$ and work out its adjoint and 
coadjoint representations, using a $2\pi L$-periodic coordinate $u$ to parametrize the circle. We refer to 
\cite{guieu2007algebre,Unterberger:2011yya} for more details on the 
Virasoro group and its cohomology.

Since the 
differentiable, real-valued second cohomology space of $G$ is three-dimensional 
\cite{Unterberger:2011yya,RogerUnter}, there are exactly 
three central extensions to be taken into account when defining $\hat G$; in other 
words, $\hat G\cong G\times\mathbb{R}^3$ as manifolds. Accordingly, elements of $\hat G$ are pairs $(f,p)$ 
belonging to $G$, supplemented by triples of real numbers $(\lambda,\mu,\nu)$. The group operation in $\hat 
G$ is
\begin{eqnarray}
&   & (f_1,p_1;\lambda_1,\mu_1,\nu_1)\cdot(f_2,p_2;\lambda_2,\mu_2,\nu_2)
=
\label{groupOp}\\
& = &
\Big(
f_1\circ f_2,
p_1+\sigma_{f_1}p_2;
\lambda_1+\lambda_2+B(f_1,f_2),
\mu_1+\mu_2+C(f_1,p_2),
\nu_1+\nu_2+D(f_1,p_1,p_2)
\Big)\,, \nonumber
\end{eqnarray}
where $\sigma$ is the action (\ref{aact}) while $B$, $C$ and $D$ are non-trivial 2-cocycles on $G$ given 
explicitly by
\begin{eqnarray}
B(f_1,f_2) & = & -\frac{1}{48\pi}\int_{S^1}\ln(f_1'\circ f_2)\extd\,\ln(f_2')\,,\\
C(f_1,p_2) & = & -\frac{1}{2\pi}\int_{S^1}p_2\cdot \extd\,\ln(f_1')\,,\\
D(f_1,p_1,p_2) & = & -\frac{1}{4\pi}\int_{S^1}p_1\cdot \extd\,(\sigma_{f_1}p_2)\,.
\end{eqnarray}
In particular, $B$ is the standard Bott--Thurston cocycle \cite{guieu2007algebre} defining the Virasoro group.

\subsubsection*{Adjoint representation and Lie brackets}

To write down an explicit formula for the coadjoint representation of the warped Virasoro group, we first 
need to work out the adjoint representation, which acts on the Lie algebra $\hat{\mathfrak{g}}$ of $\hat G$. 
As follows from the definition of $\hat G$, that algebra consists of 5-tuples $(t,p;\lambda,\mu,\nu)$ where 
$t=t(u)\frac{\partial}{\partial u}$ is a vector field on the circle, $p=p(u)$ is a function on the 
circle, and $\lambda$, $\mu$, $\nu$ are real numbers. The adjoint representation of $\hat G$, which we will 
denote as $\text{Ad}$, is then defined as
\begin{eqnarray}
&   & \text{Ad}_{(f,p_1;\lambda_1,\mu_1,\nu_1)}(t,p_2;\lambda_2,\mu_2,\nu_2)=\nonumber\\
& = &
\frac{\extd}{\extd\epsilon}\left.\left[
(f,p_1;\lambda_1,\mu_1,\nu_1)\cdot\left(e^{\epsilon t},\epsilon p_2;\epsilon 
\lambda_2,\epsilon \mu_2,\epsilon \nu_2\right)
\cdot(f,p_1;\lambda_1,\mu_1,\nu_1)^{-1}
\right]\right|_{\epsilon=0}
\label{add}
\end{eqnarray}
where $e^{\epsilon t}$ is to be understood as an infinitesimal 
diffeomorphism $e^{\epsilon t}(u)=u+\epsilon t(u)+{\cal O}(\epsilon^2)$. Given the group 
operation (\ref{groupOp}), it 
is easy to verify that the central terms $\lambda_1$, $\mu_1$ and 
$\nu_1$ play a passive role, so we may simply set 
them to zero and write $\Ad_{(f,p;\lambda,\mu,\nu)}\equiv\Ad_{(f,p)}$. Using multiple Taylor 
expansions and integrations by parts, the right-hand side of (\ref{add}) yields
\begin{eqnarray}
&   & \Ad_{(f,p_1)}(t,p_2;\lambda,\mu,\nu)=
\bigg(
\Ad_f\,t,
\sigma_fp_2+\Sigma_{\Ad_ft}\,p_1;\,
\lambda-\frac{1}{24\pi}\int\limits_0^{2\pi L}\extd u \,t(u)\{f;u\},\nonumber\\
&   &
\mu-\frac{1}{2\pi}\int p_2\cdot \extd\,\ln(f')+\frac{1}{2\pi}\int\limits_0^{2\pi L} \extd u\, t(u)
\left.\left[(p_1\circ f)''-(p_1\circ f)'\frac{f''}{f'}\right]\right|_{u},\nonumber\\
&   & \nu-\frac{1}{2\pi}\int p_1\cdot \extd\,(\sigma_fp_2)+\frac{1}{4\pi}\int\limits_0^{2\pi L}\extd u \,
t(u)
\left[(p_1\circ f)'(u)\right]^2
\bigg)\,,
\label{Ad}
\end{eqnarray}
where prime denotes differentiation with respect to $u$. Let us explain the meaning of the symbols 
appearing here:
\begin{itemize}
\item The symbol $\Ad$ on the right-hand side denotes the adjoint representation of the group $\Diff$:
\begin{equation}
\left(\Ad_ft\right)(u)
\equiv
\frac{\extd}{\extd\epsilon}\left.\left[
f\circ e^{\epsilon t}\circ f^{-1}(u)
\right]\right|_{\epsilon=0}
=
f'(f^{-1}(u))\cdot t(f^{-1}(u))\,.
\end{equation}
The far right-hand side of this equation should be seen as the component of a vector field 
$(\Ad_ft)(u)\frac{\extd}{\extd u}$. 
Equivalently,
\begin{equation}
\left(\Ad_ft\right)(f(u))
=
f'(u)\cdot t(u)\,,
\end{equation}
which is the usual transformation law of vector fields on the circle under diffeomorphisms.
\item The quantity $\{f;u\}$ is the Schwarzian derivative of the diffeomorphism $f$ evaluated at $u$:
\begin{equation}
\label{Schw}
\{f;u\}\equiv
\left.\left[\frac{f'''}{f'}-\frac{3}{2}\left(\frac{f''}{f'}\right)^2\right]\right|_{u}\,.
\end{equation}
\item The symbol $\Sigma$ denotes the differential of the action $\sigma$ of $\Diff$ on $\Cinf$. Explicitly, 
if $t$ is a vector field on the circle and if $p\in\Cinf$,
\begin{equation}
(\Sigma_t\,p)(u)
\equiv
-\frac{\extd}{\extd\epsilon}\left.\left[\left(\sigma_{e^{\epsilon t}}p\right)(u)\right]\right|_{\epsilon=0}
=
t(u)\cdot p'(u)\,.
\end{equation}
\end{itemize}

It is easily verified, upon considering an infinitesimal diffeomorphism $f$ and an infinitesimal function 
$p_1$, that the Lie brackets defined by this adjoint representation coincide with the standard brackets of a 
centrally extended warped Virasoro algebra. More precisely, upon defining the generators
\begin{equation}
T_n \equiv \left(Le^{inu/L}\frac{\partial}{\partial u},0;0,0,0\right)
\qquad
P_n \equiv \left(0,e^{inu/L};0,0,0\right)
\end{equation}
and the central charges
\begin{equation}
\label{Z1}
Z_1 \equiv \left(0,0;1,0,0\right)
\qquad
Z_2 \equiv \left(0,0;0,1,0\right)
\qquad
Z_3 \equiv \left(0,0;0,0,1\right)
\end{equation}
the Lie brackets defined by
\begin{equation}
\left[
(t_1,p_1;\lambda_1,\mu_1,\nu_1),(t_2,p_2;\lambda_2,\mu_2,\nu_2)
\right]
\equiv
-\frac{\extd}{\extd\epsilon}\left.
\left[
\Ad_{(e^{\epsilon t_1},\epsilon p_1;\epsilon \lambda_1,\epsilon 
\mu_1,\epsilon \nu_1)}(t_2,p_2;\lambda_2,\mu_2,\nu_2)
\right]
\right|_{\epsilon=0}
\end{equation}
turn out to read
\begin{eqnarray}
i[T_n,T_m] & = & (n-m)T_{n+m}+\frac{Z_1}{12}n^3\delta_{n+m,0}\\
i[T_n,P_m] & = & -mP_{n+m}-iZ_2n^2\delta_{n+m,0}\\
i[P_n,P_m] & = & Z_3n\delta_{n+m,0}\,.
\end{eqnarray}
Here we recognize the centrally extended algebra (\ref{Star}), up to the fact that the central charges $Z_i$ 
are written as operators; eventually they will be 
multiples of the identity, with coefficients $c$, $\kappa$ and $K$ corresponding to $Z_1$, $Z_2$ and 
$Z_3$ respectively.

\subsubsection*{Coadjoint representation}

The coadjoint representation of $\hat G$ is the dual of the adjoint representation, and coincides with the 
finite transformation laws of the functions $T$ and $P$ introduced in 
(\ref{eq:r26}) [i.e.~the stress tensor and the $u(1)$ current]. Explicitly, the dual $\hat{\mathfrak g}^*$ of 
the Lie algebra $\hat{\mathfrak g}$ consists of $5$-tuples
\begin{equation}
(T,P;c,\kappa,K)
\label{CoadVec}
\end{equation}
where $T=T(u)\extd u\otimes\extd u$ is a quadratic density on the circle, $P=P(u)\extd u$ is a one-form on 
the 
circle, and $c$, $\kappa$, $K$ are real numbers --- those are the values of the various central 
charges. We 
define the pairing of $\hat{\mathfrak g}^*$ with $\hat{\mathfrak g}$ by\footnote{As usual 
\cite{Witten:1987ty,guieu2007algebre}, what we call the ``dual space'' here is really the {\it smooth} dual 
space, i.e.~the space of regular distributions on the space of functions or vector fields on the circle.}
\begin{equation}
\left<
(T,P;c,\kappa,K),(t,p;\lambda,\mu,\nu)
\right>
\equiv
\frac{k}{2\pi}\int\limits_0^{2\pi L}\extd u\Big(T(u)t(u)+P(u)p(u)\Big)+c\lambda+\kappa\mu+K\nu\,,
\label{charge}
\end{equation}
so that it coincides, up to central terms, with the 
definition of surface charges (\ref{eq:r27}). Note that here $c$ is the usual Virasoro central charge, $K$ 
is the $u(1)$ level and $\kappa$ is the twist central charge appearing in (\ref{Star}). The coadjoint 
representation $\Ad^*$ of $\hat G$ is defined by
\begin{equation}
\Ad^*_{(f,p)}(T,P;c,\kappa,K)\equiv(T,P;c,\kappa,K)\circ\Ad_{(f,p)^{-1}}\,.
\nonumber
\end{equation}
Using the explicit form (\ref{Ad}) 
of the adjoint representation, one can read off the 
transformation law of each component in (\ref{CoadVec}). The result is
\begin{equation}
\Ad^*_{(f,p)}(T,P;c,\kappa,K)=\left(\Ad^*_{(f,p)}T,\Ad^*_{(f,p)}P;c,\kappa,K\right)
\end{equation}
(i.e.~the central charges are left invariant by the action of $G$), where $\Ad^*_{(f,p)}T$ and 
$\Ad^*_{(f,p)}P$ are a quadratic density and a one-form on the circle 
(respectively) whose components, evaluated at $f(u)$, are
\begin{multline}
\label{AtsfT}
 \left(\Ad^*_{(f,p)}T\right)(f(u))= 
\frac{1}{(f'(u))^2} \times\\
\times\left[
T(u)+\frac{c}{12k}\{f;u\}-P(u)(p\circ f)'(u)
-\frac{\kappa}{k}(p\circ f)''(u)+\frac{K}{2k}((p\circ f)'(u))^2
\right] 
\end{multline}
and
\begin{equation}
\left(\Ad^*_{(f,p)}P\right)(f(u))
=
\frac{1}{f'(u)}
\left[
P(u)+\frac{\kappa}{k}\frac{f''(u)}{f'(u)}-\frac{K}{k}(p\circ f)'(u)
\right]\,.
\label{AtsfJ}
\end{equation}
These are the transformation laws displayed in (\ref{tsfT})-(\ref{tsfJ}), with $\Ad^*_{(f,p)}T\equiv\tilde T$ 
and $\Ad^*_{(f,p)}P\equiv \tilde P$. They reduce to the transformations of a standard 
warped CFT \cite{Detournay:2012pc} for $\kappa=0$. In the Rindler case, however, 
$c=K=0$ and $\kappa=k$ is non-zero.

The transformation law of the function $\boost(u)$ in (\ref{3.15}) under finite asymptotic symmetry 
transformations can be worked out in a much simpler way. Indeed, it is easily verifed that the left 
formula in (\ref{3.15}) reproduces (\ref{eq:r28}) for infinitesimal transformations, and is compatible with 
the group operation (\ref{gggroup}) of the warped Virasoro group. One can also check that 
the transformation laws (\ref{3.19}) follow from (\ref{3.15}) and the definition (\ref{eq:r26}).

\subsection{Induced representations}\label{ureps}

As indicated by the imaginary vacuum values (\ref{eq:r199}) (which are actually fairly common in the world of 
warped CFT's \cite{Detournay:2012pc}), the asymptotic symmetry group is not represented 
in a unitary way in quasi-Rindler holography. Nevertheless, since the standard interpretation of symmetries 
in quantum mechanics requires unitarity \cite{Weinberg:1995I}, it is illuminating to study 
unitary representations of the warped Virasoro group. Here we classify such representations for the case of 
vanishing Kac-Moody level $K$, but 
non-vanishing twist $\kappa$. As in the case of the Euclidean, Poincar\'e or BMS 
groups, the semi-direct product structure (\ref{wVirr}) [or similarly (\ref{eq:r25})] is crucial; indeed, all 
irreducible unitary representations of 
such a group are induced \`a la Wigner \cite{Mackey,
Wigner:1939cj}. 
We refer to \cite{Cornwell}
for more details on induced representations and we mostly use the notations 
of \cite{Barnich:2014kra,Oblak:2015sea}.

\subsubsection*{A lightning review of induced representations}

The construction of induced representations of the warped Virasoro group $\hat{G}$ with vanishing $u(1)$ 
level follows the same steps as for the Poincar\'e group 
\cite{Wigner:1939cj,Wigner:1962bww} or the BMS$_3$ group \cite{Barnich:2014kra}. 
One begins 
by identifying the dual space of the 
Abelian group $\text{C}^{\infty}(S^1)$, which in the present case consists of currents $P(u)\extd u$. [The 
elements of this dual space are typically called ``momenta'', and our notation $P(u)$ is 
consistent with that terminology.] One 
then 
defines the orbit ${\cal O}_P$ and the little group $G_P$ of $P=P(u)\extd u$ as
\begin{equation}
{\cal O}_P\equiv\left\{f\cdot P\,|\,f\in\Diff\right\}
\quad\text{and}\quad
G_P\equiv\left\{f\in\Diff\,|\,f\cdot P=P\right\},
\label{OJ}
\end{equation}
where the action of $f$ on $P$ is given by (\ref{eq:3.20}). Then, given an orbit ${\cal O}_P$ and an 
irreducible, unitary 
representation ${\cal R}$ of its little group $G_P$ in some Hilbert space ${\cal E}$, the corresponding 
induced representation ${\cal T}$ acts on square-integrable wave functions $\Psi:{\cal O}_P\rightarrow{\cal 
E}:q\mapsto\Psi(q)$ according to \cite{Cornwell} 
\begin{equation}
\left({\cal T}\left[\left(f,p\right)\right]\Psi\right)(q)
\equiv
\left[\rho_{f^{-1}}(q)\right]^{1/2}
e^{i\langle q,p\rangle}{\cal R}\left[g_q^{-1}fg_{f^{-1}\cdot q}\right]\Psi\left(
f^{-1}\cdot q\right),
\label{induced}
\end{equation}
where $(f,p)$ belongs to $\Diff\ltimes\Cinf$. Let us briefly 
explain the terms of this equation:
\begin{itemize}
\item The real, positive function $\rho_f$ on ${\cal O}_P$ denotes the Radon-Nikodym derivative of the 
measure 
used to define the scalar product of wavefunctions. It is an ``anomaly'' that takes the value 
$\rho_f=1$ for all $f$ when the measure is invariant, but otherwise depends on $f$ and on the point $q$ at 
which it is evaluated. In simple cases (e.g.~the Poincar\'e group), the measure is invariant and 
$\rho_f(q)=1$ for all $f$ and all $q\in{\cal O}_P$.
\item The operator ${\cal R}\left[g_q^{-1}fg_{f^{-1}\cdot q}\right]$ is a ``Wigner rotation'': it is the 
transformation corresponding to $f$ in the space of spin degrees of freedom of the representation ${\cal T}$. 
We denote by $g_q$ the ``standard boost'' associated with $q$, that is, a group element such that 
$g_q\cdot P=q$. For scalar representations, ${\cal R}$ is trivial and one may simply forget 
about the Wigner rotation.
\end{itemize}
The classification of irreducible, unitary representations of the central extension of $\Diff\ltimes\Cinf$ 
with vanishing 
Kac-Moody level then amounts to the classification of all possible orbits (\ref{OJ}) and of all unitary 
representations of the corresponding little groups.

\subsubsection*{Induced representations of $\hat{G}$}

Our goal now is to classify all irreducible, unitary representations of the warped Virasoro 
group with vanishing Kac-Moody level, under the assumption that there exists a quasi-invariant measure on all 
the orbits (see \cite{ShavgulidzeTheOriginal} 
for the construction 
of such measures). According to the lightning review just displayed, this amounts to the classification of 
orbits, as defined in (\ref{OJ}). We start with two preliminary 
observations:
\begin{enumerate}
\item For any constant current $P(u)=P_0=\text{const}$, the little group $G_P$ consists of rigid time 
translations $f(u)=u+u_0$.
\item The charge $Q[P]$ defined as
\begin{equation}
Q[P]\equiv\frac{k}{2\pi}\int\limits_0^{2\pi L}\extd u\,P(u)
\label{QQQ}
\end{equation}
is constant along any coadjoint orbit of the warped Virasoro group, 
regardless of the values of the central charges $c$, $\kappa$ and $K$. In other words, for any 
current $P$, 
any (orientation-preserving) diffeomorphism $f$ of the circle and any function $p$, the 
zero-mode of $P$ is left invariant by the coadjoint action:
\begin{equation}
Q\left[\Ad^*_{(f,p)}P\right]
=
Q[P],
\label{QQ}
\end{equation}
where $\Ad^*_{(f,p)}P$ is given by (\ref{tsfJ}). This result holds, in particular, in the case $c=K=0$ that 
we wish to study, and corresponds physically to the fact that the average value of acceleration is invariant 
under asymptotic symmetries.
\end{enumerate}
The proof of both results is straightforward, as they can be verified by brute force. In fact, they follow 
from a stronger statement: it turns out that the orbits ${\cal O}_P$ foliate the space of currents into 
hyperplanes of constant $Q[P]$, so that {\it any} current $P(u)$ can be brought to a constant by acting with 
a 
diffeomorphism. To prove this, note that constancy of $Q[P]$ implies that that constant, if it exists, 
coincides with the zero-mode 
$P_0$ of $P(u)$. The question thus boils down to whether or not there exists an orientation-preserving 
diffeomorphism $f$ 
such that
\begin{equation}
\left.\left(f\cdot P\right)\right|_{f(u)}
\stackrel{\text{(\ref{3.15})}}{=}
\frac{1}{f'(u)}
\left[
P(u)+\frac{\kappa}{k}\frac{f''(u)}{f'(u)}
\right]
\stackrel{!}{=}
P_0\,.
\end{equation}
This condition is equivalent to an inhomogeneous first-order 
differential equation for $1/f'$, whose solution is
\begin{equation}
\frac{1}{f'(u)}
=
A\,\exp\left[\frac{k}{\kappa}\int\limits_0^u\extd v\,P(v)\right]
-
\frac{k}{\kappa}P_0
\,\int\limits_0^u\extd t\,\exp\left[\frac{k}{\kappa}\int\limits_t^u\extd v\,P(v)\right]
\end{equation}
where $A$ is a real parameter. Since $u$ is assumed to be $2\pi L$-periodic, this function must be $2\pi 
L$-periodic as well. This selects a unique solution (and such a solution exists for any value of 
$P_0$), meaning that, for any $P(u)$, there always exists a diffeomorphism $f$ of the circle such that 
$f\cdot P$ be a constant\footnote{An alternative way to 
prove the same result is to recall the modified 
Sugawara construction (\ref{Suga}), by which a coadjoint orbit of the Virasoro group with 
non-negative energy is associated with each orbit ${\cal O}_P$. Since the only Virasoro orbits with 
non-negative energy are orbits of constants, we know that there always exists a diffeomorphism $f$ that 
brings 
a given Sugawara stress tensor (\ref{Suga}) into a constant, which in turn brings the corresponding 
current 
$P$ to a constant.}; furthermore, that diffeomorphism is uniquely specified by $P(u)$ up to a rigid 
time translation.

This proves that all orbits ${\cal O}_P$ are hyperplanes specified by the value of the charge 
(\ref{QQQ}), in accordance with the fact that $P_0$ is a Casimir operator in (\ref{Star}). One can then apply 
the usual machinery 
of induced representations to the warped Virasoro group, and compute, for instance, the associated characters 
along the lines of \cite{Oblak:2015sea}; however, as the interpretation of these characters in the 
present context is unclear, we refrain from displaying them.

\subsubsection*{Physical properties of unitary representations}

We have just seen that unitary representations of the warped Virasoro group [with vanishing $u(1)$ level] can 
be classified according to the possible values of the Casimir
operator $P_0$ in (\ref{Star}). Accordingly, the orbits 
(\ref{OJ}) are affine hyperplanes with constant zero-modes embedded in the space of currents 
$P(u)$. In particular, each orbit contains exactly one constant representative. The physical meaning of this 
statement is that any (generally time-dependent) 
acceleration $a(u)$ can be brought to a constant, $a_0$, by using a suitable reparametrization of time that 
preserves $2\pi L$-periodicity. Furthermore, $a_0$ coincides with the Fourier zero-mode of $a(u)$. Note that, 
with the requirement of $2\pi L$-periodicity, it is no longer true that any time-dependent acceleration 
$a(u)$ can be mapped on $\tilde{a}=0$ because the diffeomorphisms defined by (\ref{s7q}) generally do not 
preserve that requirement. 

Having classified the orbits, we know, in principle, the irreducible unitary 
representations of the warped Virasoro group at vanishing Kac-Moody level. In three space\-time dimensions, 
these representations describe the Hilbert space of metric fluctuations around the background specified by 
the 
orbit ${\cal O}_P$ and the representation ${\cal R}$ 
\cite{Giombi:2008vd,Garbarz:2014kaa,Barnich:2014kra,Barnich:2015mui}, as follows from the fact 
that the phase space coincides with the coadjoint representation 
of the asymptotic symmetry group. For instance, an 
induced 
representation of $\hat G$ specified by an orbit ${\cal O}_P$ and the trivial 
representation of the little group gives the Hilbert space of metric fluctuations around the 
background
\begin{equation}
\extd s^2 = -2P(u)\,r\,\extd u^2 - 2\extd u\extd r + \extd x^2.
\end{equation}
Here $u$ is still understood as a $2\pi L$-periodic coordinate; in particular, the solution at $a=0$ is {\it 
not} Minkowski space\-time because of that identification.

Note that, in any unitary representation of the type just described, the eigenvalues of the Hamiltonian $T_0$ 
are unbounded from below (and from above). There is thus a trade-off 
between unitarity and 
boundedness of energy: if we insist that the representation be unitary, then it is a (direct integral of) 
induced representation(s), and energy is unbounded both from below and from above; conversely, if we insist 
that energy be bounded from below, then the asymptotic symmetry group cannot act unitarily on the Hilbert 
space of the putative dual theory. This property has actually been observed in representations of the 
Galilean Conformal Algebra in two 
dimensions, $\mathfrak{gca}_2$, and its higher-spin extensions \cite{Bagchi:2009ca,Afshar:2013vka}. 
Indeed, when $T_0$ is interpreted as the Hamiltonian, demanding that energy be bounded from below amounts to 
considering highest-weight representations of the symmetry algebras (\ref{eq:r30}) or (\ref{bmsSuga}), the 
highest weight being the lowest eigenvalue of $T_0$ in the space of the representation. This representation 
is easily seen to be non-unitary when the central charge of the mixed commutator is non-zero. We stress, 
however, that in the more common interpretation of the warped Virasoro group \cite{Detournay:2012pc} where 
$P_0$ plays the role of the Hamiltonian, there is no such conflict between unitarity and boundedness of 
energy.

\section{Rindler thermodynamics}\label{app:B}

We have not found a holographic setup that leads to Rindler thermodynamics, but it is still of interest in 
its 
own right to consider it. In this appendix we describe Rindler thermodynamics in a non-holographic context 
and 
show that it recovers the near horizon thermodynamics of BTZ black holes and flat space cosmologies, in the 
sense that temperatures and entropies agree with each other. If a consistent version of Rindler 
holography exists, it should reproduce the results of this appendix (see also \cite{Laflamme:1987ec}).

The main change as compared to the main text is that the periodicities in the Euclidean coordinates are no 
longer given by \eqref{eq:r33}, but instead by
\eq{
(t_{\textrm{\tiny E}},\,y)\sim(t_{\textrm{\tiny E}}+\beta,\,y-i\beta\boost)\sim(t_{\textrm{\tiny 
E}},\,y+\tilde 
L)
}{eq:rindlerwahnsinn}
where $\tilde L$ is now the periodicity of the spatial coordinate $y$ and $\beta$, $\boost$ coincide with the 
quasi-Rindler parameters. So in Rindler thermodynamics we do not identify retarded time periodically, which 
is the key difference to quasi-Rindler thermodynamics.

\subsection{Rindler horizon and temperature}

The Euclidean metric \eqref{eq:r14} has a center at $\rho=0$, corresponding to the Rindler horizon in 
Lorentzian signature. In order for this center to be smooth, the Euclidean time $t_{\textrm{\tiny E}}\sim 
t_{\textrm{\tiny E}}+\beta$ has to have a periodicity $\beta=2\pi/a$. Interpreting the inverse of this 
periodicity as temperature yields the  expected Unruh temperature \cite{Unruh:1976db}
\eq{
T = \frac{a}{2\pi}\,.
}{eq:r31}
The same result is obtained from surface gravity. Note that the Unruh temperature \eqref{eq:r31} is 
independent of the boost parameter $\f$.

There is yet another way to determine the Unruh temperature, namely starting from rotating (non-extremal) BTZ 
and taking the near horizon limit. The rotating BTZ metric \cite{Banados:1992wn} 
\eq{
 \mathrm{d}s^2 = -\frac{(r^2 - r_+^2)(r^2 - r_-^2)}{\ell^2r^2}\, \mathrm{d}t^2 + \frac{\ell^2r^2}{(r^2 - 
r_+^2)(r^2 - r_-^2)} \,\mathrm{d}r^2 + r^2 \,\big( \mathrm{d}\varphi - \frac{r_+r_-}{\ell r^2} \,\mathrm{d}t 
\big)^2
}{eq:r40}
leads to a Hawking temperature
\eq{
T_H = \frac{r_+^2-r_-^2}{2\pi r_+\ell^2}\,.
}{eq:r41}
Now take the near horizon limit by defining $\rho = (r^2 - r_+^2)/(2r_+)$ and dropping higher order terms in 
$\rho$, which gives
\eq{
\extd s^2 = -2a\rho\,\extd t^2 + \frac{\extd \rho^2}{2a\rho} + \big(\extd x + \f\,\extd t\big)^2
}{eq:r42}
with
\eq{
a = \frac{\hat r_+^2 - \hat r_-^2/\ell^2}{\hat r_+},\qquad \f = -\hat r_-, \qquad x = \hat r_+ \ell^2 \varphi,
}{eq:r43}
where $\hat r_+ = r_+ / \ell^2$ and $\hat r_- = r_- / \ell$.
Note that in the limit of infinite AdS radius, $\ell\to\infty$, we keep fixed the rescaled parameters $\hat 
r_\pm$, and the coordinate $x$ decompactifies. [We then recompactify by imposing \eqref{eq:rindlerwahnsinn}.] 
The Hawking temperature $T_H$ can be rewritten as
\eq{
T_H = \frac{\hat r_+^2-\hat r_-^2/\ell^2}{2\pi \hat r_+} = \frac{a}{2\pi} 
}{eq:r44}
and thus coincides with the Unruh temperature \eqref{eq:r31}. Besides verifying this expected result, the 
calculation above provides expressions for the Rindler parameters $a$ and $\f$ in terms of BTZ parameters 
$r_\pm$, which can be useful for other consistency checks as well. 

Essentially the same conclusion holds for flat space cosmologies \cite{Cornalba:2002fi}, 
whose metric reads 
\eq{
\extd s^2 = r_+^2\big(1-\tfrac{r_0^2}{r^2}\big)\,\extd t^2 - \frac{\extd 
r^2}{r_+^2\big(1-\tfrac{r_0^2}{r^2}\big)} + r^2\,\big(\extd\varphi - \tfrac{r_+ r_0}{r^2}\,\extd t\big)^2\,.
}{eq:r46}
In the near horizon approximation, $r^2=r_0^2+2r_0\rho$, we recover the line-element \eqref{eq:r42} with
\eq{
a = -\frac{r_+^2}{r_0} \qquad \f = -r_+ \qquad x = r_0\varphi\,.
}{eq:r47}
The cosmological temperature $T=r_+^2/(2\pi r_0)$ again coincides with the Unruh temperature \eqref{eq:r31}, 
up to a sign. This sign is explained by inner horizon black hole mechanics 
\cite{Castro:2012av}. 

The fact that Hawking/cosmological temperatures coincide with the Rindler temperature is not surprising but 
follows from kinematics. What is less clear is whether or not extensive quantities like free energy or 
entropy 
coincide as well. We calculate these quantities in the next two subsections.

\subsection{Rindler free energy}

Since we have no Rindler boundary conditions we do not know what the correct on-shell action is, as we 
have no 
way of checking the variational principle. However, since the zero-mode solutions of quasi-Rindler 
holography 
coincide with the zero-mode solutions used in Rindler thermodynamics it is plausible that the action 
\eqref{eq:r17} can be used again. We base our discussion of free energy and entropy on this assumption.

Evaluating the full action \eqref{eq:r17} on-shell and multiplying it by temperature $T=\beta^{-1}$ yields 
the free energy,
\eq{
F = -\frac{T}{8\pi G_N}\,\int\limits_0^{\tilde L}\extd y \int\limits_0^\beta\extd t_{\textrm{\tiny 
E}}\sqrt{\ga}\,K\big|_{\rho\to\infty}\,.
}{eq:r36}
The quantity $\tilde L$ denotes the range of the coordinate $y$ and physically corresponds to the horizon 
area. If $\tilde L$ tends to infinity we simply define densities of extensive quantities like free energy or 
entropy by dividing all such expressions by $\tilde L$.
Insertion of the boosted Rindler metric \eqref{eq:r14} into the general expression for free energy \eqref{eq:r36} yields
\eq{
F = -\frac{a \tilde L}{8\pi G_N} = -\frac{T \tilde L}{4G_N}\,.
}{eq:r22}

It is worthwhile mentioning that Rindler free energy \eqref{eq:r22} does not coincide with the corresponding 
BTZ or FSC free energy. Using the identifications \eqref{eq:r43} and \eqref{eq:r47} we find in both cases 
$F_{\textrm{\tiny BTZ}} = F_{\textrm{\tiny FSC}} = - T \tilde L/(8G_N)$, which differs by a factor $1/2$ from 
the Rindler result \eqref{eq:r22}. Nevertheless, as we shall demonstrate below, the corresponding entropies 
do coincide.

\subsection{Rindler entropy}

Our result for free energy \eqref{eq:r22} implies that Rindler entropy obeys the Bekenstein--Hawking area law,
\eq{
S = -\frac{\extd F}{\extd T} = \frac{\tilde L}{4G_N}.
}{eq:r35}
Note that entropy does not go to zero at arbitrarily small temperature. However, in that regime one should not trust the Rindler approximation since the $T\to 0$ limit is more adequately modelled by extremal horizons rather than non-extremal ones. 

The result for entropy \eqref{eq:r35} can be obtained within the first order formulation as well. Applying 
the flat space results of \cite{Gary:2014ppa} 
to the present case yields
\eq{
S = \frac{k}{2\pi}\,\int\limits_0^\beta\extd u\int\limits_0^{\tilde L}\extd x\, \langle A_u A_x\rangle  = 
\frac{k}{2\pi}\,\tilde L\,\beta\,\langle \frak a_u \frak a_x\rangle = k\,\tilde L\,.
}{eq:r37a}
Relating the Chern--Simons level with the inverse Newton constant, $k=1/(4G_N)$ then reproduces precisely the 
Bekenstein--Hawking area law \eqref{eq:r35}.

Interestingly, Rindler entropy \eqref{eq:r35} also follows from near horizon BTZ entropy. The latter is given by
\eq{
S_{\textrm{\tiny BTZ}} = \frac{2\pi r_+}{4 G_N} = \frac{2\pi\hat r_+ \ell^2}{4 G_N} = \frac{\tilde L}{4G_N}\,.
}{eq:r45}
In the last equality we identified the length of the $x$-interval using the last relation \eqref{eq:r43} 
together with $\varphi\sim\varphi+2\pi$. Thus, the near horizon BTZ entropy coincides with the Rindler 
entropy, which provides another consistency check on the correctness of our result. 

The same conclusions hold for the entropy of flat space cosmologies,
\eq{
S_{\textrm{\tiny FSC}} = \frac{2\pi r_0}{4 G_N} = \frac{\tilde L}{4G_N}\,.
}{eq:r48}
In the last equality we identified the length of the $x$-interval using the last relation \eqref{eq:r47} 
together with $\varphi\sim\varphi+2\pi$. 

The results above confirm that entropy is a near-horizon property, whereas free energy and the conserved charges are a property of the global spacetime.

\bibliographystyle{fullsort}
\addcontentsline{toc}{section}{References}

\begin{thebibliography}{100}

\bibitem{Staruszkiewicz:1963zza}
A.~Staruszkiewicz, ``{Gravitation Theory in Three-Dimensional Space},'' {\em
  Acta Phys. Polon.} {\bf 24} (1963)
735--740.

S.~Deser, R.~Jackiw, and G.~'t~Hooft, ``Three-dimensional einstein gravity:
  Dynamics of flat space,'' {\em Ann. Phys.} {\bf 152} (1984)
220.

S.~Deser and R.~Jackiw, ``Three-dimensional cosmological gravity: Dynamics of
  constant curvature,'' {\em Annals Phys.} {\bf 153} (1984)
405--416.

\bibitem{Banados:1992wn}
M.~Ba\~nados, C.~Teitelboim, and J.~Zanelli, ``The black hole in
  three-dimensional space-time,'' {\em Phys. Rev. Lett.} {\bf 69} (1992)
  1849--1851,
\href{http://www.arXiv.org/abs/hep-th/9204099}{{\tt hep-th/9204099}}.

M.~Ba\~nados, M.~Henneaux, C.~Teitelboim, and J.~Zanelli, ``Geometry of the
  (2+1) black hole,'' {\em Phys. Rev.} {\bf D48} (1993) 1506--1525,
\href{http://www.arXiv.org/abs/gr-qc/9302012}{{\tt gr-qc/9302012}}.

\bibitem{Cornalba:2002fi}
L.~Cornalba and M.~S. Costa, ``{A New cosmological scenario in string
  theory},'' {\em Phys.Rev.} {\bf D66} (2002) 066001,
\href{http://www.arXiv.org/abs/hep-th/0203031}{{\tt hep-th/0203031}}.
%
``{Time dependent orbifolds and string
  cosmology},'' {\em Fortsch.Phys.} {\bf 52} (2004) 145--199,
\href{http://www.arXiv.org/abs/hep-th/0310099}{{\tt hep-th/0310099}}.

\bibitem{Brown:1986nw}
J.~D. Brown and M.~Henneaux, ``{Central Charges in the Canonical Realization of
  Asymptotic Symmetries: An Example from Three-Dimensional Gravity},'' {\em
  Commun. Math. Phys.} {\bf 104} (1986)
207--226.

\bibitem{Henneaux:2002wm}
M.~Henneaux, C.~Martinez, R.~Troncoso, and J.~Zanelli, ``{Black holes and
  asymptotics of 2+1 gravity coupled to a scalar field},'' {\em Phys. Rev.}
  {\bf D65} (2002) 104007,
\href{http://www.arXiv.org/abs/hep-th/0201170}{{\tt hep-th/0201170}}.

\bibitem{Grumiller:2008es}
D.~Grumiller and N.~Johansson, ``{Consistent boundary conditions for
  cosmological topologically massive gravity at the chiral point},'' {\em Int.
  J. Mod. Phys.} {\bf D17} (2009) 2367--2372,
\href{http://www.arXiv.org/abs/0808.2575}{{\tt 0808.2575}}.

M.~Henneaux, C.~Martinez, and R.~Troncoso, ``{Asymptotically anti-de Sitter
  spacetimes in topologically massive gravity},'' {\em Phys. Rev.} {\bf D79}
  (2009) 081502R,
\href{http://www.arXiv.org/abs/0901.2874}{{\tt 0901.2874}}.

\bibitem{Compere:2013bya}
G.~Comp{\`e}re, W.~Song, and A.~Strominger, ``{New Boundary Conditions for
  AdS$_3$},'' {\em JHEP} {\bf 1305} (2013) 152,
\href{http://www.arXiv.org/abs/1303.2662}{{\tt 1303.2662}}.

C.~Troessaert, ``{Enhanced asymptotic symmetry algebra of $AdS$$_{3}$},'' {\em
  JHEP} {\bf 1308} (2013) 044,
\href{http://www.arXiv.org/abs/1303.3296}{{\tt 1303.3296}}.

\bibitem{Barnich:2006av}
G.~Barnich and G.~Comp{\`e}re, ``{Classical central extension for asymptotic
  symmetries at null infinity in three spacetime dimensions},'' {\em
  Class.Quant.Grav.} {\bf 24} (2007) F15--F23,
\href{http://www.arXiv.org/abs/gr-qc/0610130}{{\tt gr-qc/0610130}}.

\bibitem{Ashtekar:1996cd}
A.~Ashtekar, J.~Bicak, and B.~G. Schmidt, ``{Asymptotic structure of symmetry
  reduced general relativity},'' {\em Phys.Rev.} {\bf D55} (1997) 669--686,
\href{http://www.arXiv.org/abs/gr-qc/9608042}{{\tt gr-qc/9608042}}.

\bibitem{Bagchi:2009my}
A.~Bagchi and R.~Gopakumar, ``{Galilean Conformal Algebras and AdS/CFT},'' {\em
  JHEP} {\bf 0907} (2009) 037,
\href{http://www.arXiv.org/abs/0902.1385}{{\tt 0902.1385}}.

\bibitem{Bagchi:2009pe}
A.~Bagchi, R.~Gopakumar, I.~Mandal, and A.~Miwa, ``{GCA in 2d},'' {\em JHEP}
  {\bf 1008} (2010) 004,
\href{http://www.arXiv.org/abs/0912.1090}{{\tt 0912.1090}}.

\bibitem{Bagchi:2010zz}
A.~Bagchi, ``{Correspondence between Asymptotically Flat Spacetimes and
  Nonrelativistic Conformal Field Theories},'' {\em Phys.Rev.Lett.} {\bf 105}
  (2010) 171601,
\href{http://www.arXiv.org/abs/1006.3354}{{\tt 1006.3354}}.

\bibitem{Bagchi:2012yk}
A.~Bagchi, S.~Detournay, and D.~Grumiller, ``{Flat-Space Chiral Gravity},''
  {\em Phys.Rev.Lett.} {\bf 109} (2012) 151301,
\href{http://www.arXiv.org/abs/1208.1658}{{\tt 1208.1658}}.

\bibitem{Barnich:2012xq}
G.~Barnich, ``{Entropy of three-dimensional asymptotically flat cosmological
  solutions},'' {\em JHEP} {\bf 1210} (2012) 095,
\href{http://www.arXiv.org/abs/1208.4371}{{\tt 1208.4371}}.

A.~Bagchi, S.~Detournay, R.~Fareghbal, and J.~Simon, ``{Holography of 3d Flat
  Cosmological Horizons},'' {\em Phys. Rev. Lett.} {\bf 110} (2013) 141302,
\href{http://www.arXiv.org/abs/1208.4372}{{\tt 1208.4372}}.

\bibitem{Bagchi:2013lma}
A.~Bagchi, S.~Detournay, D.~Grumiller, and J.~Simon, ``{Cosmic Evolution from
  Phase Transition of Three-Dimensional Flat Space},'' {\em Phys.Rev.Lett.}
  {\bf 111} (2013) 181301,
\href{http://www.arXiv.org/abs/1305.2919}{{\tt 1305.2919}}.

\bibitem{Detournay:2014fva}
S.~Detournay, D.~Grumiller, F.~Sch{\"o}ller, and J.~Simon, ``{Variational
  principle and 1-point functions in 3-dimensional flat space Einstein
  gravity},'' {\em Phys.Rev.} {\bf D89} (2014) 084061,
\href{http://www.arXiv.org/abs/1402.3687}{{\tt 1402.3687}}.

\bibitem{Afshar:2013vka}
H.~Afshar, A.~Bagchi, R.~Fareghbal, D.~Grumiller, and J.~Rosseel, ``{Higher
  spin theory in 3-dimensional flat space},'' {\em Phys.Rev.Lett.} {\bf 111}
  (2013) 121603,
\href{http://www.arXiv.org/abs/1307.4768}{{\tt 1307.4768}}.

H.~A. Gonz\'alez, J.~Matulich, M.~Pino, and R.~Troncoso, ``{Asymptotically flat
  spacetimes in three-dimensional higher spin gravity},'' {\em JHEP} {\bf 1309}
  (2013) 016,
\href{http://www.arXiv.org/abs/1307.5651}{{\tt 1307.5651}}.

\bibitem{Barnich:2014kra}
G.~Barnich and B.~Oblak, ``{Notes on the BMS group in three dimensions: I.
  Induced representations},'' {\em JHEP} {\bf 1406} (2014) 129,
\href{http://www.arXiv.org/abs/1403.5803}{{\tt 1403.5803}}.
%
``{Notes on the BMS group in three dimensions: II.
  Coadjoint representation},'' {\em JHEP} {\bf 1503} (2015) 033,
\href{http://www.arXiv.org/abs/1502.00010}{{\tt 1502.00010}}.

\bibitem{Oblak:2015sea}
B.~Oblak, ``{Characters of the BMS Group in Three Dimensions},'' {\em Commun.
  Math. Phys.} {\bf 340} (2015), no.~1, 413--432,
\href{http://www.arXiv.org/abs/1502.03108}{{\tt 1502.03108}}.

\bibitem{Barnich:2015mui}
G.~Barnich, H.~A. Gonz\'alez, A.~Maloney, and B.~Oblak, ``{One-loop partition
  function of three-dimensional flat gravity},'' {\em JHEP} {\bf 1504} (2015)
  178,
\href{http://www.arXiv.org/abs/1502.06185}{{\tt 1502.06185}}.

\bibitem{Bagchi:2014iea}
A.~Bagchi, R.~Basu, D.~Grumiller, and M.~Riegler, ``{Entanglement entropy in
  Galilean conformal field theories and flat holography},'' {\em
  Phys.Rev.Lett.} {\bf 114} (2015), no.~11, 111602,
\href{http://www.arXiv.org/abs/1410.4089}{{\tt 1410.4089}}.

S.~M. Hosseini and A.~Veliz-Osorio, ``{Gravitational anomalies, entanglement
  entropy, and flat-space holography},''
\href{http://www.arXiv.org/abs/1507.06625}{{\tt 1507.06625}}.

R.~Basu and M.~Riegler,
  ``Wilson Lines and Holographic Entanglement Entropy in Galilean Conformal Field Theories,''
\href{http://www.arXiv.org/abs/1511.08662}{{\tt 1511.08662}}.

\bibitem{Barnich:2012aw}
G.~Barnich, A.~Gomberoff, and H.~A. Gonz\'alez, ``{The Flat limit of three
  dimensional asymptotically anti-de Sitter spacetimes},'' {\em Phys.Rev.} {\bf
  D86} (2012) 024020,
\href{http://www.arXiv.org/abs/1204.3288}{{\tt 1204.3288}}.

A.~Bagchi and R.~Fareghbal, ``{BMS/GCA Redux: Towards Flatspace Holography from
  Non-Relativistic Symmetries},'' {\em JHEP} {\bf 1210} (2012) 092,
\href{http://www.arXiv.org/abs/1203.5795}{{\tt 1203.5795}}.

A.~Bagchi, ``{Tensionless Strings and Galilean Conformal Algebra},'' {\em JHEP}
  {\bf 1305} (2013) 141,
\href{http://www.arXiv.org/abs/1303.0291}{{\tt 1303.0291}}.

R.~Caldeira~Costa, ``{Aspects of the zero $\Lambda$ limit in the AdS/CFT
  correspondence},'' {\em Phys.Rev.} {\bf D90} (2014), no.~10, 104018,
\href{http://www.arXiv.org/abs/1311.7339}{{\tt 1311.7339}}.

R.~Fareghbal and A.~Naseh, ``{Flat-Space Energy-Momentum Tensor from BMS/GCA
  Correspondence},'' {\em JHEP} {\bf 1403} (2014) 005,
\href{http://www.arXiv.org/abs/1312.2109}{{\tt 1312.2109}}.

C.~Krishnan, A.~Raju, and S.~Roy, ``{A Grassmann path from $AdS_3$ to flat
  space},'' {\em JHEP} {\bf 1403} (2014) 036,
\href{http://www.arXiv.org/abs/1312.2941}{{\tt 1312.2941}}.

A.~Bagchi and R.~Basu, ``{3D Flat Holography: Entropy and Logarithmic
  Corrections},'' {\em JHEP} {\bf 1403} (2014) 020,
\href{http://www.arXiv.org/abs/1312.5748}{{\tt 1312.5748}}.

G.~Barnich, L.~Donnay, J.~Matulich, and R.~Troncoso, ``{Asymptotic symmetries
  and dynamics of three-dimensional flat supergravity},'' {\em JHEP} {\bf 1408}
  (2014) 071,
\href{http://www.arXiv.org/abs/1407.4275}{{\tt 1407.4275}}.

M.~Riegler, ``{Flat space limit of higher-spin Cardy formula},'' {\em
  Phys.Rev.} {\bf D91} (2015), no.~2, 024044,
\href{http://www.arXiv.org/abs/1408.6931}{{\tt 1408.6931}}.

R.~Fareghbal and A.~Naseh, ``{Aspects of Flat/CCFT Correspondence},'' {\em
  Class.Quant.Grav.} {\bf 32} (2015), no.~13, 135013,
\href{http://www.arXiv.org/abs/1408.6932}{{\tt 1408.6932}}.

R.~Fareghbal and S.~M. Hosseini, ``{Holography of 3D Asymptotically Flat Black
  Holes},'' {\em Phys. Rev.} {\bf D91} (2015), no.~8, 084025,
\href{http://www.arXiv.org/abs/1412.2569}{{\tt 1412.2569}}.

\bibitem{Gary:2014ppa}
M.~Gary, D.~Grumiller, M.~Riegler, and J.~Rosseel, ``{Flat space (higher spin)
  gravity with chemical potentials},'' {\em JHEP} {\bf 1501} (2015) 152,
\href{http://www.arXiv.org/abs/1411.3728}{{\tt 1411.3728}}.

J.~Matulich, A.~Perez, D.~Tempo, and R.~Troncoso, ``{Higher spin extension of
  cosmological spacetimes in 3D: asymptotically flat behaviour with chemical
  potentials and thermodynamics},''
\href{http://www.arXiv.org/abs/1412.1464}{{\tt 1412.1464}}.

\bibitem{Bagchi:2015wna}
A.~Bagchi, D.~Grumiller, and W.~Merbis, ``{Stress tensor correlators in
  three-dimensional gravity},''
\href{http://www.arXiv.org/abs/1507.05620}{{\tt 1507.05620}}.

\bibitem{Rindler:1966zz}
W.~Rindler, ``{Kruskal Space and the Uniformly Accelerated Frame},'' {\em
  Am.J.Phys.} {\bf 34} (1966)
1174.

\bibitem{Fulling:1972md}
S.~A. Fulling, ``{Nonuniqueness of canonical field quantization in Riemannian
  space-time},'' {\em Phys.Rev.} {\bf D7} (1973)
2850--2862.

P.~Davies, ``{Scalar particle production in Schwarzschild and Rindler
  metrics},'' {\em J.Phys.} {\bf A8} (1975)
609--616.

\bibitem{Unruh:1976db}
W.~G. Unruh, ``Notes on black hole evaporation,'' {\em Phys. Rev.} {\bf D14}
  (1976)
870.

W.~G. Unruh and R.~M. Wald, ``{What happens when an accelerating observer
  detects a Rindler particle},'' {\em Phys.Rev.} {\bf D29} (1984)
1047--1056.

\bibitem{Laflamme:1987ec}
R.~Laflamme, ``{Entropy of a Rindler Wedge},'' {\em Phys.Lett.} {\bf B196}
  (1987)
449--450.

\bibitem{Lowe:1994ah}
D.~A. Lowe and A.~Strominger, ``{Strings near a Rindler or black hole
  horizon},'' {\em Phys.Rev.} {\bf D51} (1995) 1793--1799,
\href{http://www.arXiv.org/abs/hep-th/9410215}{{\tt hep-th/9410215}}.

S.~Deser and O.~Levin, ``{Accelerated detectors and temperature in (anti)-de
  Sitter spaces},'' {\em Class. Quant. Grav.} {\bf 14} (1997) L163--L168,
\href{http://www.arXiv.org/abs/gr-qc/9706018}{{\tt gr-qc/9706018}}.
%
``{Equivalence of Hawking and Unruh temperatures through
  flat space embeddings},'' {\em Class. Quant. Grav.} {\bf 15} (1998) L85--L87,
\href{http://www.arXiv.org/abs/hep-th/9806223}{{\tt hep-th/9806223}}.
%
``{Mapping Hawking into Unruh thermal properties},''
  {\em Phys. Rev.} {\bf D59} (1999) 064004,
\href{http://www.arXiv.org/abs/hep-th/9809159}{{\tt hep-th/9809159}}.

S.~Carlip, ``Entropy from conformal field theory at killing horizons,'' {\em
  Class. Quant. Grav.} {\bf 16} (1999) 3327--3348,
\href{http://www.arXiv.org/abs/gr-qc/9906126}{{\tt gr-qc/9906126}}.

T.~Padmanabhan, ``{Thermodynamics and / of horizons: A Comparison of
  Schwarzschild, Rindler and de Sitter space-times},'' {\em Mod.Phys.Lett.}
  {\bf A17} (2002) 923--942,
\href{http://www.arXiv.org/abs/gr-qc/0202078}{{\tt gr-qc/0202078}}.

V.~Moretti and N.~Pinamonti, ``{Holography and SL(2,R) symmetry in 2-D Rindler
  space-time},'' {\em J.Math.Phys.} {\bf 45} (2004) 230,
\href{http://www.arXiv.org/abs/hep-th/0304111}{{\tt hep-th/0304111}}.

D.~Marolf, D.~Mini\'c, and S.~F. Ross, ``{Notes on space-time thermodynamics and
  the observer dependence of entropy},'' {\em Phys.Rev.} {\bf D69} (2004)
  064006,
\href{http://www.arXiv.org/abs/hep-th/0310022}{{\tt hep-th/0310022}}.

A.~J. Amsel, D.~Marolf, and A.~Virmani, ``{The Physical Process First Law for
  Bifurcate Killing Horizons},'' {\em Phys.Rev.} {\bf D77} (2008) 024011,
\href{http://www.arXiv.org/abs/0708.2738}{{\tt 0708.2738}}.

H.~Casini, ``{Relative entropy and the Bekenstein bound},'' {\em
  Class.Quant.Grav.} {\bf 25} (2008) 205021,
\href{http://www.arXiv.org/abs/0804.2182}{{\tt 0804.2182}}.

\bibitem{Chung:2010ge}
H.~Chung, ``{Asymptotic Symmetries of Rindler Space at the Horizon and Null
  Infinity},'' {\em Phys.Rev.} {\bf D82} (2010) 044019,
\href{http://www.arXiv.org/abs/1005.0820}{{\tt 1005.0820}}.

\bibitem{Grumiller:2010bz}
D.~Grumiller, ``{Model for gravity at large distances},'' {\em Phys.Rev.Lett.}
  {\bf 105} (2010) 211303, \href{http://www.arXiv.org/abs/1011.3625}{{\tt
  1011.3625}}.

S.~Carloni, D.~Grumiller, and F.~Preis, ``{Solar system constraints on Rindler
  acceleration},'' {\em Phys.Rev.} {\bf D83} (2011) 124024,
  \href{http://www.arXiv.org/abs/1103.0274}{{\tt 1103.0274}}.

D.~Grumiller, M.~Irakleidou, I.~Lovrekovi\'c, and R.~McNees, ``{Conformal gravity
  holography in four dimensions},'' {\em Phys.Rev.Lett.} {\bf 112} (2014)
  111102,
\href{http://www.arXiv.org/abs/1310.0819}{{\tt 1310.0819}}.

\bibitem{Parikh:2011aa}
M.~Parikh, P.~Samantray, and E.~Verlinde, ``{Rotating Rindler-AdS Space},''
  {\em Phys.Rev.} {\bf D86} (2012) 024005,
\href{http://www.arXiv.org/abs/1112.3433}{{\tt 1112.3433}}.

B.~Czech, J.~L. Karczmarek, F.~Nogueira, and M.~Van~Raamsdonk, ``{Rindler
  Quantum Gravity},'' {\em Class.Quant.Grav.} {\bf 29} (2012) 235025,
\href{http://www.arXiv.org/abs/1206.1323}{{\tt 1206.1323}}.

M.~Parikh and P.~Samantray, ``{Rindler-AdS/CFT},''
\href{http://www.arXiv.org/abs/1211.7370}{{\tt 1211.7370}}.

T.~G. Mertens, H.~Verschelde, and V.~I. Zakharov, ``{Random Walks in Rindler
  Spacetime and String Theory at the Tip of the Cigar},'' {\em JHEP} {\bf 1403}
  (2014) 086,
\href{http://www.arXiv.org/abs/1307.3491}{{\tt 1307.3491}}.

E.~Halyo, ``{Rindler Energy is Wald Entropy},''
\href{http://www.arXiv.org/abs/1403.2333}{{\tt 1403.2333}}.
%
``{On the Holographic Nature Of Rindler Energy},''
\href{http://www.arXiv.org/abs/1406.5763}{{\tt 1406.5763}}.
%
``{Black Holes as Conformal Field Theories on Horizons},''
\href{http://www.arXiv.org/abs/1502.01979}{{\tt 1502.01979}}.

R.~Fareghbal and A.~Naseh, ``{Rindler/Contracted-CFT Correspondence},'' {\em
  JHEP} {\bf 1406} (2014) 134,
\href{http://www.arXiv.org/abs/1404.3937}{{\tt 1404.3937}}.

\bibitem{Achucarro:1987vz}
A.~Achucarro and P.~K. Townsend, ``A {C}hern-{S}imons action for
  three-dimensional {A}nti-de {S}itter supergravity theories,'' {\em Phys.
  Lett.} {\bf B180} (1986)
89.

E.~Witten, ``(2+1)-dimensional gravity as an exactly soluble system,'' {\em
  Nucl. Phys.} {\bf B311} (1988)
46.

\bibitem{Barnich:2013yka}
G.~Barnich and H.~A. Gonz\'alez, ``{Dual dynamics of three dimensional
  asymptotically flat Einstein gravity at null infinity},'' {\em JHEP} {\bf
  1305} (2013) 016,
\href{http://www.arXiv.org/abs/1303.1075}{{\tt 1303.1075}}.

\bibitem{Afshar:2013bla}
H.~R. Afshar, ``{Flat/AdS boundary conditions in three dimensional conformal
  gravity},'' {\em JHEP} {\bf 1310} (2013) 027,
\href{http://www.arXiv.org/abs/1307.4855}{{\tt 1307.4855}}.

\bibitem{York:1972sj}
J.~W. York, Jr., ``Role of conformal three geometry in the dynamics of
  gravitation,'' {\em Phys. Rev. Lett.} {\bf 28} (1972)
1082--1085.

G.~W. Gibbons and S.~W. Hawking, ``Action integrals and partition functions in
  quantum gravity,'' {\em Phys. Rev.} {\bf D15} (1977)
2752--2756.

\bibitem{Gary:2012ms}
M.~Gary, D.~Grumiller, and R.~Rashkov, ``{Towards non-AdS holography in
  3-dimensional higher spin gravity},'' {\em JHEP} {\bf 1203} (2012) 022,
\href{http://www.arXiv.org/abs/1201.0013}{{\tt 1201.0013}}.

H.~Afshar, M.~Gary, D.~Grumiller, R.~Rashkov, and M.~Riegler, ``{Non-AdS
  holography in 3-dimensional higher spin gravity - General recipe and
  example},'' {\em JHEP} {\bf 1211} (2012) 099,
\href{http://www.arXiv.org/abs/1209.2860}{{\tt 1209.2860}}.

\bibitem{Maldacena:1997re}
J.~M. Maldacena, ``{The large $N$ limit of superconformal field theories and
  supergravity},'' {\em Adv. Theor. Math. Phys.} {\bf 2} (1998) 231--252,
\href{http://www.arXiv.org/abs/hep-th/9711200}{{\tt hep-th/9711200}}.

\bibitem{Detournay:2012pc}
S.~Detournay, T.~Hartman, and D.~M. Hofman, ``{Warped Conformal Field
  Theory},'' {\em Phys.Rev.} {\bf D86} (2012) 124018,
\href{http://www.arXiv.org/abs/1210.0539}{{\tt 1210.0539}}.

\bibitem{Compere:2008cv}
G.~Comp\`ere and S.~Detournay, ``{Semi-classical central charge in
  topologically massive gravity},'' {\em Class. Quant. Grav.} {\bf 26} (2009)
  012001,
\href{http://www.arXiv.org/abs/0808.1911}{{\tt 0808.1911}}.

\bibitem{Compere:2009zj}
G.~Comp\`ere and S.~Detournay, ``{Boundary conditions for spacelike and
  timelike warped AdS$_3$ spaces in topologically massive gravity},'' {\em
  JHEP} {\bf 08} (2009) 092,
\href{http://www.arXiv.org/abs/0906.1243}{{\tt 0906.1243}}.

M.~Blagojevi\'c and B.~Cvetkovi\'c, ``{Asymptotic structure of topologically
  massive gravity in spacelike stretched AdS sector},'' {\em JHEP} {\bf 09}
  (2009) 006,
\href{http://www.arXiv.org/abs/0907.0950}{{\tt 0907.0950}}.

M.~Henneaux, C.~Martinez, and R.~Troncoso, ``{Asymptotically warped anti-de
  Sitter spacetimes in topologically massive gravity},'' {\em Phys. Rev.} {\bf
  D84} (2011) 124016,
\href{http://www.arXiv.org/abs/1108.2841}{{\tt 1108.2841}}.

\bibitem{Bertin:2012qw}
M.~Bertin, S.~Ertl, H.~Ghorbani, D.~Grumiller, N.~Johansson, and
  D.~Vassilevich, ``{Lobachevsky holography in conformal Chern-Simons
  gravity},'' {\em JHEP} {\bf 1306} (2013) 015,
\href{http://www.arXiv.org/abs/1212.3335}{{\tt 1212.3335}}.

\bibitem{Afshar:2011yh}
H.~Afshar, B.~Cvetkovi\'c, S.~Ertl, D.~Grumiller, and N.~Johansson, ``{Holograms
  of Conformal Chern-Simons Gravity},'' {\em Phys.Rev.} {\bf D84} (2011)
  041502(R), \href{http://www.arXiv.org/abs/1106.6299}{{\tt 1106.6299}}.
%
``{Conformal
  Chern-Simons holography - lock, stock and barrel},'' {\em Phys.Rev.} {\bf
  D85} (2012) 064033,
\href{http://www.arXiv.org/abs/1110.5644}{{\tt 1110.5644}}.

\bibitem{Hofman:2014loa}
D.~M. Hofman and B.~Rollier, ``{Warped Conformal Field Theory as Lower Spin
  Gravity},'' {\em Nucl. Phys.} {\bf B897} (2015) 1--38,
\href{http://www.arXiv.org/abs/1411.0672}{{\tt 1411.0672}}.

\bibitem{Hofman:2011zj}
D.~M. Hofman and A.~Strominger, ``{Chiral Scale and Conformal Invariance in 2D
  Quantum Field Theory},'' {\em Phys. Rev. Lett.} {\bf 107} (2011) 161601,
\href{http://www.arXiv.org/abs/1107.2917}{{\tt 1107.2917}}.

\bibitem{Banados:1994tn}
M.~Ba\~nados, ``{Global charges in Chern-Simons field theory and the (2+1) black
  hole},'' {\em Phys. Rev.} {\bf D52} (1996) 5816--5825,
\href{http://www.arXiv.org/abs/hep-th/9405171}{{\tt hep-th/9405171}}.

\bibitem{Afshar:2014rwa}
H.~Afshar, A.~Bagchi, S.~Detournay, D.~Grumiller, S.~Prohazka, and M.~Riegler,
  ``{Holographic Chern-Simons Theories},'' {\em Lect.Notes Phys.} {\bf 892}
  (2015) 311--329,
\href{http://www.arXiv.org/abs/1404.1919}{{\tt 1404.1919}}.

\bibitem{Regge:1974zd}
T.~Regge and C.~Teitelboim, ``{Role of Surface Integrals in the Hamiltonian
  Formulation of General Relativity},'' {\em Annals Phys.} {\bf 88} (1974)
286.

G.~Barnich and F.~Brandt, ``{Covariant theory of asymptotic symmetries,
  conservation laws and central charges},'' {\em Nucl. Phys.} {\bf B633} (2002)
  3--82,
\href{http://www.arXiv.org/abs/hep-th/0111246}{{\tt hep-th/0111246}}.

G.~Barnich and G.~Comp\`ere, ``{Surface charge algebra in gauge theories and
  thermodynamic integrability},'' {\em J. Math. Phys.} {\bf 49} (2008) 042901,
\href{http://www.arXiv.org/abs/0708.2378}{{\tt 0708.2378}}.

\bibitem{Garbarz:2014kaa}
A.~Garbarz and M.~Leston, ``{Classification of Boundary Gravitons in AdS$_3$
  Gravity},'' {\em JHEP} {\bf 05} (2014) 141,
\href{http://www.arXiv.org/abs/1403.3367}{{\tt 1403.3367}}.

\bibitem{1403.3835}
G.~Barnich and B.~Oblak, ``{Holographic positive energy theorems in
  three-dimensional gravity},''
\href{http://www.arXiv.org/abs/1403.3835}{{\tt 1403.3835}}.

\bibitem{Compere:2015knw}
G.~Comp\`ere, P.-J. Mao, A.~Seraj, and S.~Sheikh-Jabbari, ``{Symplectic and
  Killing Symmetries of AdS$_3$ Gravity: Holographic vs Boundary Gravitons},''
\href{http://www.arXiv.org/abs/1511.06079}{{\tt 1511.06079}}.

\bibitem{Unterberger:2011yya}
J.~Unterberger and C.~Roger, {\em The Schr{\"o}dinger--Virasoro Algebra:
  Mathematical structure and dynamical Schr{\"o}dinger symmetries}.
\newblock Theoretical and Mathematical Physics. Springer Berlin Heidelberg,
2011.
\newblock

\bibitem{Shaghoulian:2015dwa}
E.~Shaghoulian, ``{A Cardy formula for holographic hyperscaling-violating
  theories},'' {\em JHEP} {\bf 11} (2015) 081,
\href{http://www.arXiv.org/abs/1504.02094}{{\tt 1504.02094}}.

\bibitem{Donnay:2015abr}
L.~Donnay, G.~Giribet, H.~A. Gonz\'alez, and M.~Pino, ``{Super-translations and
  super-rotations at the horizon},''
\href{http://www.arXiv.org/abs/1511.08687}{{\tt 1511.08687}}.

\bibitem{Hawking:2015qqa}
S.~W. Hawking, ``{The Information Paradox for Black Holes},''
\newblock 2015.
\newblock
\href{http://www.arXiv.org/abs/1509.01147}{{\tt 1509.01147}}.
\newblock

M.~J. Perry, ``{Black Hole Memory},''
\newblock talk delivered at the Nordic Institute for Theoretical Physics, 2015.

\bibitem{Hawking:2016msc} 
  S.~W.~Hawking, M.~J.~Perry and A.~Strominger,
  ``Soft Hair on Black Holes,''
\href{http://www.arXiv.org/abs/1601.00921}{{\tt 1601.00921}}.

\bibitem{Blau:2015nee}
M.~Blau and M.~O'Loughlin, ``{Horizon Shells and BMS-like Soldering
  Transformations},''
\href{http://www.arXiv.org/abs/1512.02858}{{\tt 1512.02858}}.

\bibitem{Ayon-Beato:2015xsz}
E.~Ay{\'o}n-Beato and G.~Vel{\'a}zquez-Rodr\'i­guez, ``{On the Residual
  Symmetries of the Gravitational Field},''
\href{http://www.arXiv.org/abs/1511.07461}{{\tt 1511.07461}}.

\bibitem{Deser:1982vy}
S.~Deser, R.~Jackiw, and S.~Templeton, ``Three-dimensional massive gauge
  theories,'' {\em Phys. Rev. Lett.} {\bf 48} (1982)
975--978.
%
``Topologically massive gauge
  theories,'' {\em Ann. Phys.} {\bf 140} (1982)
372--411.

\bibitem{guieu2007algebre}
L.~Guieu and C.~Roger, {\em L'alg{\`e}bre et le groupe de Virasoro: aspects
  g{\'e}om{\'e}triques et alg{\'e}briques, g{\'e}n{\'e}ralisations}.
\newblock les Publications CRM, 2007.

\bibitem{RogerUnter}
C.~{Roger} and J.~{Unterberger}, ``{The Schr{\"o}dinger--Virasoro Lie
  group and algebra: from geometry to representation theory},'' 
  \href{http://www.arXiv.org/abs/math-ph/0601050}{{\tt math-ph/0601050}}.

\bibitem{Witten:1987ty}
E.~Witten, ``{Coadjoint Orbits of the Virasoro Group},'' {\em Commun. Math.
  Phys.} {\bf 114} (1988)
1.

\bibitem{Weinberg:1995I}
S.~Weinberg, {\em The Quantum Theory of Fields}, vol.~I.
\newblock Cambridge University Press, 1995.

\bibitem{Mackey}
G.~Mackey, ``Infinite dimensional group representations,'' {\em Bull. Amer.
  Math. Soc.} {\bf 69} (1963) 628--686.
%
{\em Induced representations of groups and quantum mechanics}.
\newblock Publicazioni della Classe di Scienze della Scuola Normale Superiore
  di Pisa. W. A. Benjamin, 1968.
%
{\em Infinite Dimensional Group Representations and Their
  Applications}.
\newblock Forschungsinstitut f{\"u}r Mathematik, ETH, 1971.

\bibitem{Wigner:1939cj}
E.~P. Wigner, ``On unitary representations of the inhomogeneous {L}orentz
  group,'' {\em Annals Math.} {\bf 40} (1939) 149--204, Reprinted in: {\em Nucl.~Phys.~Proc.~Suppl.} {\bf 6} (1989) 9.

\bibitem{Cornwell}
J.~Cornwell, {\em Group theory in physics}.
\newblock Techniques of physics. Academic Press, 1984.

A.~Barut and R.~Raczka, {\em Theory of Group Representations and Applications}.
\newblock World Scientific, 1986.

\bibitem{Wigner:1962bww}
E.~P. Wigner, ``Unitary representations of the inhomogeneous {L}orentz group
  including reflections,'' in {\em Group theoretical concepts and methods in
  elementary particle physics, Lectures of the Istanbul summer school of
  theoretical physics (1962)}, F.~G\"ursey, ed., pp.~37--80.
\newblock Gordon and Breach, New York, 1964.

\bibitem{ShavgulidzeTheOriginal}
E.~Shavgulidze, ``{A measure that is quasi-invariant with respect to the action
  of a group of diffeomorphisms of a finite-dimensional manifold},'' {\em Dokl.
  Akad. Nauk SSSR} {\bf 303} (1988), no.~4, 811--814.
%
``{An example of a measure quasi-invariant under the action of
  the diffeomorphism group of the circle},'' {\em {Funkts. Anal. Prilozh.}}
  {\bf 12} (1978), no.~3, 55--60.
%
``{Mesures quasi-invariantes sur les groupes de
  diff\'eomorphismes des vari\'et\'es riemaniennes},'' {\em C. R. Acad. Sci.
  Paris} {\bf 321} (1995) 229--232.
%
``{Quasiinvariant measures on groups of diffeomorphisms.},''
  in {\em {Loop spaces and groups of diffeomorphisms. Collected papers}},
  pp.~181--202 (1997); translation from tr. mat. inst. steklova 217, 189--208.
\newblock Moscow: MAIK Nauka/Interperiodica Publishing, 1997.

\bibitem{Giombi:2008vd}
S.~Giombi, A.~Maloney, and X.~Yin, ``{One-loop Partition Functions of 3D
  Gravity},'' {\em JHEP} {\bf 0808} (2008) 007,
  \href{http://www.arXiv.org/abs/0804.1773}{{\tt 0804.1773}}.

\bibitem{Bagchi:2009ca}
A.~Bagchi and I.~Mandal, ``{On Representations and Correlation Functions of
  Galilean Conformal Algebras},'' {\em Phys.Lett.} {\bf B675} (2009) 393--397,
\href{http://www.arXiv.org/abs/0903.4524}{{\tt 0903.4524}}.

\bibitem{Castro:2012av}
A.~Castro and M.~J. Rodriguez, ``{Universal properties and the first law of
  black hole inner mechanics},'' {\em Phys.Rev.} {\bf D86} (2012) 024008,
\href{http://www.arXiv.org/abs/1204.1284}{{\tt 1204.1284}}.

S.~Detournay, ``{Inner Mechanics of 3d Black Holes},'' {\em Phys.Rev.Lett.}
  {\bf 109} (2012) 031101,
\href{http://www.arXiv.org/abs/1204.6088}{{\tt 1204.6088}}.

\end{thebibliography}

\providecommand{\href}[2]{#2}\begingroup\raggedright\endgroup

\end{document}